# SRF Cavity Fabrication and Materials

*W. Singer*[1]

Deutsches Elektronen-Synchrotron DESY, Hamburg, Germany

**Abstract**

The technological and metallurgical requirements of material for high-gradient superconducting cavities are described. High-purity niobium, as the preferred metal for the fabrication of superconducting accelerating cavities, should meet exact specifications. The content of interstitial impurities such as oxygen, nitrogen, and carbon must be below 10μg/g. The hydrogen content should be kept below 2μg/g to prevent degradation of the *Q*-value under certain cool-down conditions. The material should be free of flaws (foreign material inclusions or cracks and laminations) that can initiate a thermal breakdown. Defects may be detected by quality control methods such as eddy current scanning and identified by a number of special methods. Conventional and alternative cavity fabrication methods are reviewed. Conventionally, niobium cavities are fabricated from sheet niobium by the formation of half-cells by deep drawing, followed by trim machining and Electron-Beam Welding (EBW). The welding of half-cells is a delicate procedure, requiring intermediate cleaning steps and a careful choice of weld parameters to achieve full penetration of the joints. The equator welds are particularly critical. A challenge for a welded construction is the tight mechanical and electrical tolerances. These can be maintained by a combination of mechanical and radio-frequency measurements on half-cells and by careful tracking of weld shrinkage. The established procedure is suitable for large series production. The main aspects of quality assurance management are mentioned. Another cavity fabrication approach is slicing discs from the ingot and producing cavities by deep drawing and EBW. Accelerating gradients at the level of 35–45 MV·m$^{-1}$ can be achieved by applying Electropolishing (EP) treatment. Furthermore, the single-crystal option (grain boundary free) is promising. It seems that in this case, high performance can be achieved by a simplified treatment procedure. Fabrication of the accelerating structure from a seamless pipe as a cost-effective alternative is briefly described. This technology has yielded good results in single-cell cavities and is already available for multi-cell structures.

*Keywords*: superconducting cavities, fabrication, materials, quality assurance, accelerators.

## 1   Introduction

Niobium, having the highest critical temperature and critical magnetic field of all pure metals (critical temperature 9.3 K; superheating field of ~240 mT) has for many years now been the preferred metal for the fabrication of superconducting RF cavities [1–8]. This is because niobium is chemically inert (at room temperature, the surface is covered by a protecting pentoxide layer), it can easily be machined and deep drawn, and it is available on the market, in the required amounts, in bulk and sheet material form. In the past three decades, improvements of Electron-Beam Melting (EBM) and purification techniques have made good progress [9, 10]. The purity of industrially produced niobium has steadily increased due to the reduction of concentrations of metallic impurities such as Ta and W and of the interstitially dissolved elements carbon, oxygen, and nitrogen. The application of ultra-high-vacuum technology additionally increases purity. Several companies worldwide are currently in a position to produce tonnes of niobium for high-gradient cavities.

---

[1] waldemar.singer@desy.de

A few European companies already have about 30 years' experience in mechanical fabrication of superconducting cavities. New companies in the USA and Asia are joining this community, especially in the context of anticipating the construction of an International Linear Collider (ILC) that would require about 20 000 superconducting cavities. The new trend is not only mechanically to produce cavities in the industry, but also to ask industry to perform cavity surface treatment and provide laboratories with cavities ready for installation in accelerating cryo-modules.

The main aspects of niobium production as a base material and the manufacturing of superconducting resonators from this material are described below.

## 2 Niobium as a superconducting material for cavity fabrication

### 2.1 From ore to semi-finished product

Niobium and tantalum always occur in association with one another in nature [11–14] (Table 1).

The main mineral, pyrochlore, is mostly processed by primarily physical processing technology to obtain niobium oxide, with a concentration from 55% to about 60%. The most important reserves are calcium niobate from Brazil (about 80%) and Canada (about 10%), and zinc or tin slags containing niobium and tantalum from Zaire, Nigeria, and Russia, as well as stibiotantalite from West Australia.

The second most important source of Nb-producing ore is niobite(columbite)–tantalite [(Fe, Mn)(Nb,Ta)$_2$O$_6$]. When tantalum significantly exceeds niobium by weight, it is tantalite; and when niobium significantly exceeds tantalum by weight, it is columbite. As a mineral with a ratio of Nb$_2$O$_5$:Ta$_2$O$_5$ from 10:1 to 13:1, columbite occurs in Brazil, Nigeria, and Australia, as well as other countries of Central Africa. Usually, niobium is recovered when the ores are processed for tantalum.

The world's largest niobium deposit, located in Araxá, Brazil, is owned by Companhia Brasileira de Metalurgia e Mineração (CBMM). The reserves are enough to supply current world demand for about 500 years, about 460 million tons. The weathered ore mine contents are between 2.5 and 3.0% Nb$_2$O$_5$. The mining is carried out by open-pit extraction. The ore is crushed and magnetite is magnetically separated from the pyrochlore. By chemical processes, the ore is concentrated with regard to its Nb content (50–60% of Nb$_2$O$_5$).

**Table 1:** The most important niobium minerals

| Mineral | Composition | Percentage of Nb$_2$O$_5$ + Ta$_2$O$_5$ | Percentage of Nb$_2$O$_5$ | Percentage of Ta$_2$O$_5$ |
|---|---|---|---|---|
| Pyrochlore | NaCaNb$_2$O$_6$F | 38–73 | 26–73 | 0.2–22 |
| Niobite (columbite) | (Fe,Mn)(Nb,Ta)$_2$O$_6$ | 75–81 | 47–78 | 0.1–34 |
| Tantalite | (Fe,Mn)(Ta,Nb)$_2$O$_6$ | 81–86 | 2–27 | 53–84 |

Niobium ores are mainly processed into concentrates by opening up with hydrofluoric acid. Then the niobium and tantalum oxides are separated from one another. The method used for Ta and

Nb separation (liquid–liquid extraction or solvent extraction) is based on their relative solubility in two different immiscible liquids. The best method to produce raw material on an industrial scale is liquid–liquid extraction using methyl isobutyl ketone (MIBK: $C_6H_{12}O$). Niobium is recovered as niobium oxide $Nb_2O_5$ via neutralization of the niobium fluoride complex with ammonia to form the hydroxide, followed by calcination to the oxide. Generally, the tantalum values in solution are converted into tantalum oxide ($Ta_2O_5$). Filtration of the liquid mixture and further processing via solvent extraction using MIBK produces highly purified solutions of tantalum and niobium. The tantalum content in Nb oxides separated by liquid–liquid extraction is at a level below 500 µg/g (100 µg/g is reachable). Classical routes from niobium oxide to metal (see Fig. 1 [10]) consist of carbothermic reduction of $Nb_2O_5$ or aluminothermic reduction according to the equations

$$Nb_2O_5 + 7C \rightarrow 2NbC + 5CO,$$
$$5NbC + Nb_2O_5 \rightarrow 7Nb + 5CO,$$
$$3Nb_2O_5 + 10Al \rightarrow 5Al_2O_3 + 6Nb,$$

followed by Electron-Beam Melting (EBM).

The production of high-grade niobium with a small Ta concentration can be performed via the sodium reduction of purified $K_2NbF_7$:

$$K_2NbF_7 + 5Na \rightarrow Nb + 2KF + 5NaF.$$

An optional route for niobium fabrication is powder metallurgy. Niobium can be mechanically pulverized using cooling from 950°C in a hydrogen atmosphere. After the hydriding treatment, the metals are crushed, ground, sieved and dehydrided in a vacuum. The powder can then be mechanically pressed to compacts and used for EBM.

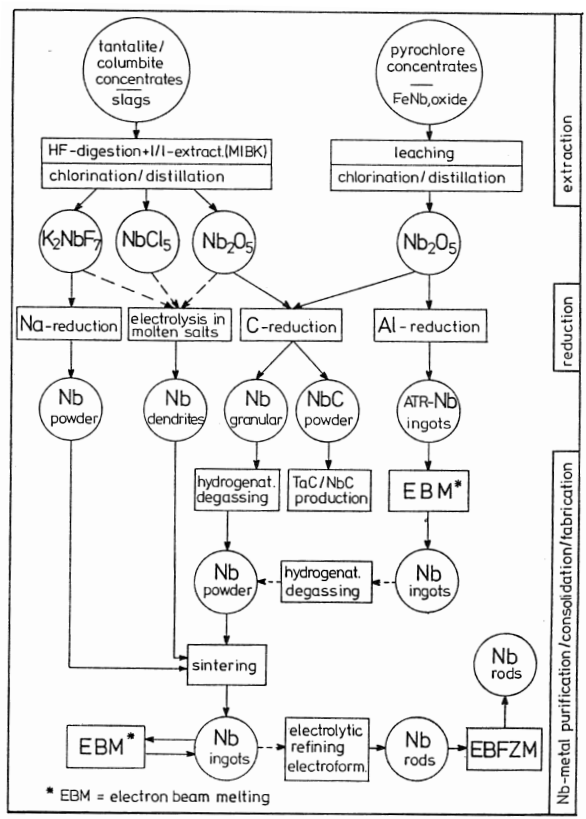

**Fig. 1:** An overview of the niobium production and purification processes [10]

## 2.2 Purification

### 2.2.1 Refining by EBM

As a result of increasing demand for refractory metals in the past few decades, the electron-beam furnace has been developed into a reliable, efficient apparatus for melting and purification [10, 15–18]. Evaporation of most impurities by EBM in a vacuum of typically better than $3 \times 10^{-4}$ mbar for the first melt and $2 \times 10^{-6}$ mbar for the last melt is very effective for niobium because of its high melting point (2468°C). Therefore, it is possible to produce tons of extremely pure niobium industrially.

The principle of EBM equipment, which is now routinely available in the industry, is shown in Fig. 2 [10].

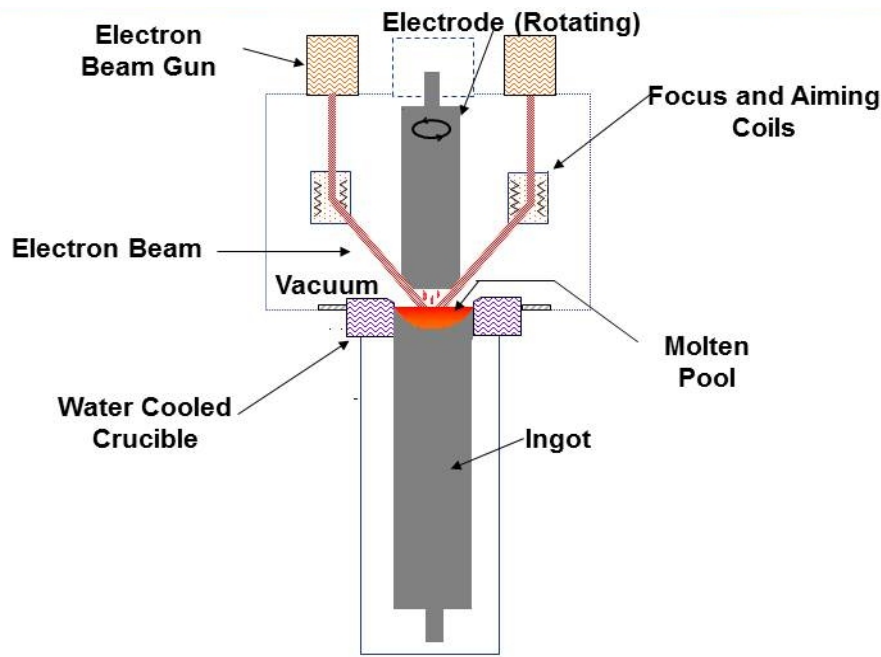

**Fig. 2:** A schematic of the EBM equipment [17]

As the ingot is melting, molten metal globules drop into a pool on the ingot, which is contained in a water-cooled copper cylinder (crucible). For the first melt, the powder, the granulate and the solid materials are pressed to form an electrode that is fed in horizontally. The material is fed in vertically in the subsequent melts. Additional purification occurs by repeatedly re-melting under vacuum. In part, the energy of the electron beam is used to melt the ingot and in part it is used to maintain the pool of metal liquid. During melting, all gases and impurities that have a melting temperature lower than that of niobium are evaporating. The ingot is continuously guided through the crucible. The rate of guidance is co-ordinated with the rate of the material melting to ensure complete melting of the feed material and proper refining. The melting temperature is a compromise between maximization of purification and minimization of material losses by evaporation.

Melted ingots sometimes demonstrate a non-homogeneous distribution of impurities from top to bottom. The skin of the ingot contains more impurities than the inside. Machining away the skin and cutting away a fraction of the ingot bottom are recommended for a purer final product.

The furnace design known as the e-beam cold hearth furnace could be an effective alternative to the method described above. This design provides maximum exposure of the melt to the vacuum environment by melting in a rectangular hearth and at the same time casting a slab shape. Such a shape is more suitable for rolling than the round ingot.

The contents of the interstitial impurities O, N, and C that significantly influence the properties of the niobium can be reduced down to a level below 10 μg/g; an H content close to 1 μg/g is reachable. This content achieved by melting should be maintained during fabrication and treatment. Four to six melting steps are generally necessary to reach the Residual Resistivity Ratio (RRR) = 300 level, with a few μg/g of interstitial impurities; an RRR of up to 500 can ultimately be achieved (Fig. 3 [18]).

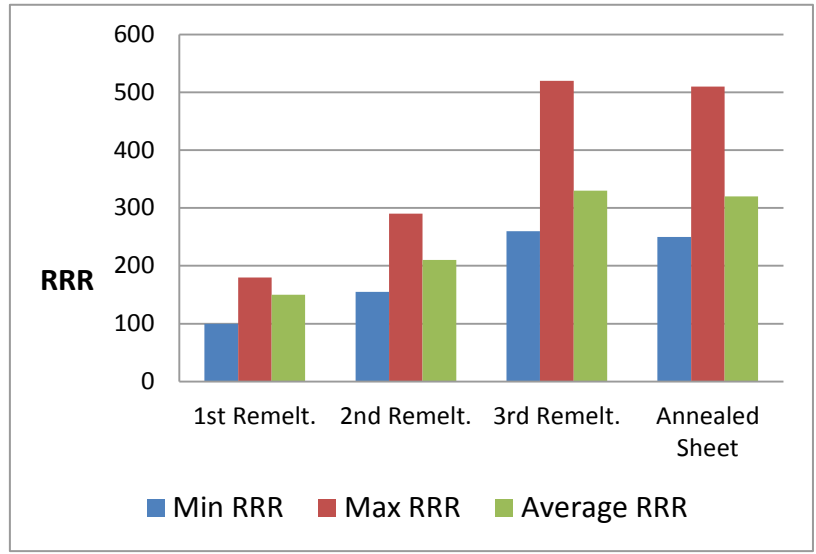

**Fig. 3:** The RRR values from the first, second and third re-melted ingot and annealed sheet [18]

Complicated processes of evaporation, degassing, solvation, decomposition, and diffusion take place during EB melting of the ingot (Fig. 4 [10]). It can be seen that for $H_2$ and $N_2$, thermodynamic equilibrium between the gas phase and the interstitially dissolved H and N atoms is established according to the reactions $H_2$ (g) ↔ 2 H (in Nb) and $N_2$ (g) ↔ 2 N (in Nb). For oxygen, a balance between the uptake of $O_2$ on the one hand and $H_2O$ and oxide evaporation as NbO or $NbO_2$ on the other is achieved. Unlike oxygen, carbon desorbs only in the form of CO.

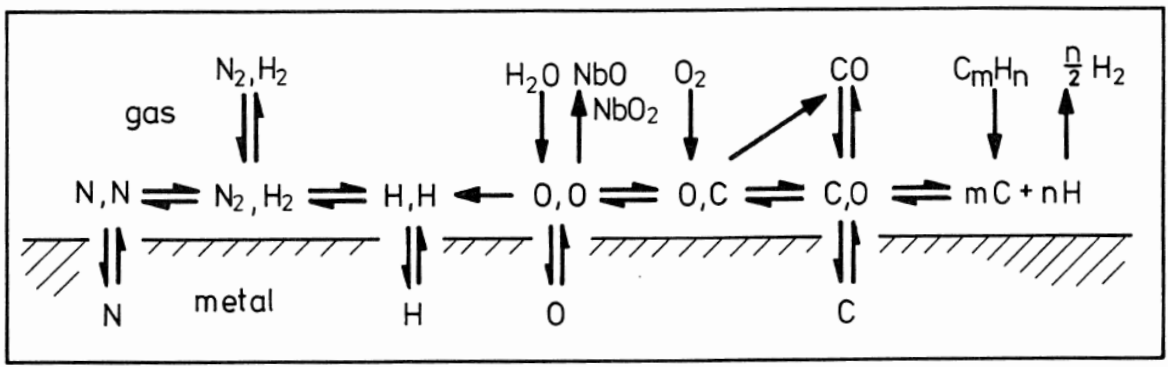

**Fig. 4:** A schematic of the gassing and degassing reactions between niobium and residual gases in vacuum at high temperatures [10].

### 2.2.2 *Purification by electrolysis in molten salts*

Fundamental investigations of the electrodeposition of Nb from LiF–NaF–KF melts containing $K_2NbF_7$, at $T$ = 700–800°C, have shown that this electrolyte is suitable for material refining (electrolytic refining) [9]. Niobium becomes extremely pure in relation to metallic elements; for example, the tantalum content can be reduced by this treatment from 400 to 3.5 μg/g (see Fig. 5).

Subsequent experiments have shown that similar results can be achieved using niobium fluoride with potassium fluoride (10–20%) and sodium fluoride (5–15%), together with an equimolecular mixture of chlorides (NaCl–KCl) for the remainder [19]. This rather expensive method was applied in the former Soviet Union in the 1980s for the industrial production of niobium, as a refining step before EBM. An RRR purity level of about 1000 was achieved in Nb sheets that were usable for cavity production.

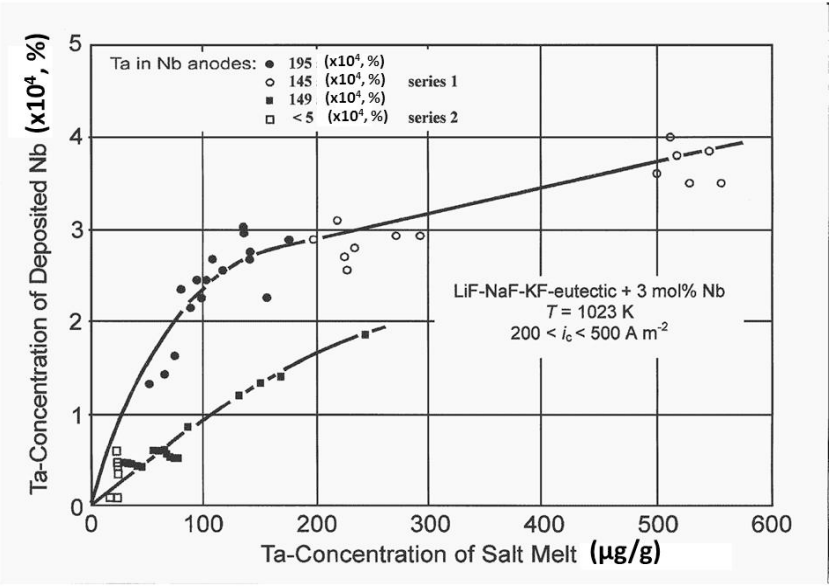

**Fig. 5:** The purification of niobium by electrolysis in molten salts [9]

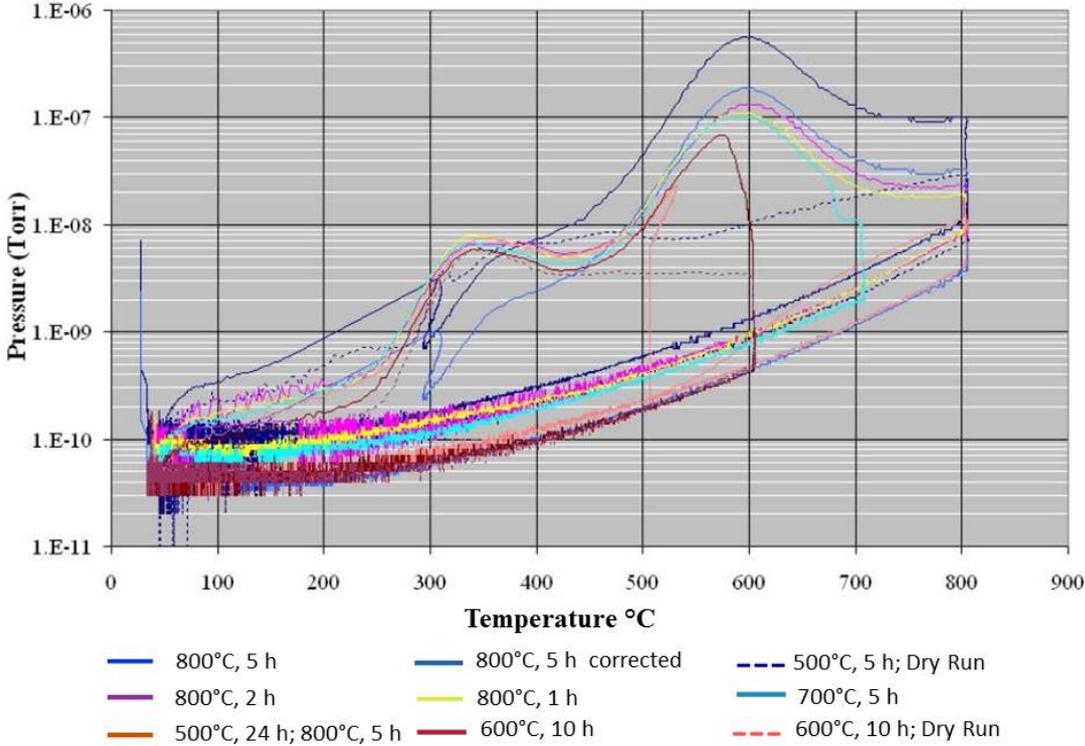

**Fig. 6:** The behaviour of the hydrogen partial pressure during annealing of niobium [21]

## 2.3 Post-purification

In some cavity surface treatment processes, it is very difficult to completely avoid contamination by interstitials. For example, contamination by hydrogen is almost inevitable during Electropolishing (EP) or Centrifugal Barrel Polishing (CBP). Furthermore, additional purification of niobium increases the expectations of better performance. The possible purification of already mechanically produced cavities is described below.

### 2.3.1 Hydrogen degassing

Hydrogen degassing can be achieved by common annealing. A plot of the hydrogen partial pressure versus temperature normally shows two peaks around 350°C and 600–650°C. It can be assumed that the peak at 350°C is caused by surface-trapped hydrogen, while the other peak indicates the bulk of the hydrogen contamination. There are some differences of opinion concerning the optimal hydrogen degassing temperatures. Annealing at 600°C for 10 h has been proposed in the literature [20]. Investigation [21] (see Fig. 6) shows that heat treatment at 800°C for 5 h will result in a larger depletion of hydrogen from the material than 600°C for 10 h. Annealing at 800°C for 2 h is applied for European X-ray Free Electron Laser (XFEL) cavities [22], which in our opinion is rather optimal and allows a hydrogen level of <1 µg/g to be reached. Annealing at 800°C has some advantages: the degassing of the oven is rather fast, and the niobium becomes completely recrystallized at this temperature. In addition, the lattice presents fewer defects after such annealing. The improved lattice is not so sensitive to further hydrogen contamination.

### 2.3.2 Solid-state gettering

One needs a very high temperature and very low partial pressures in order to degas the oxygen and nitrogen [10]. Such annealing does not make much sense for cavity application. During recrystallization annealing at 750–800°C and with a conventional vacuum of ~$10^{-6}$ mbar, some surface contamination by O and N takes place [10]. The contamination depth can be estimated on the basis of the diffusion law [23] and, as shown in Fig. 7, its penetration is about 150–200 µm.

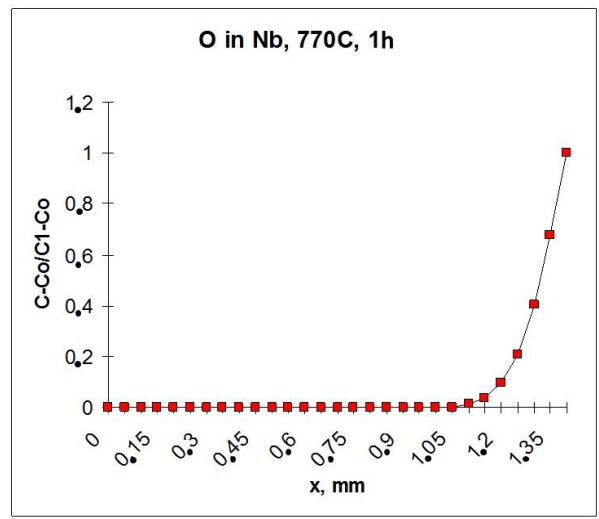

**Fig. 7:** The calculated penetration depth of oxygen into Nb during recrystallization annealing [7]

This contamination can be removed by purification heat treatment, often called solid-state gettering (see, e.g., Ref. [24]). The getter metal Me (usually Ti) is vapour deposited on the surface of the niobium at high temperature. The bonding enthalpies of this metal to the interstitial impurities, such as oxygen, nitrogen, or carbon, should be higher than that of Nb. The creation of the $Me_xO$, $Me_xN$ and Me$_x$C compounds reduces the concentration of the interstitial impurities on the surface of the Nb. Moreover, the Me protects against the absorption of the residual gas from the furnace environment. On the other hand, the high temperature intensifies the diffusion of the interstitial impurities from the inside to the surface, and as a result also allows purification of the bulk of the niobium. This procedure has another positive effect for cavity performance: the annealing itself, at high temperature, additionally homogenizes the niobium (dissolves small segregations of different types; for example, residues of oxides, clusters of foreign materials, etc.). The temperature and duration of the purification annealing depend on the evaporation rate of the getter material and the diffusion rate of the impurities. This technique is, in principle, capable of improving the RRR by a factor of 10.

Several metals, such as Ti, Y, Zr, and Hf, that have higher bonding enthalpies with oxygen, nitrogen, and carbon than niobium, can be used successfully for niobium post-purification [1, 2, 24]. Titanium has been applied for solid-state gettering of FLASH cavities at DESY, with annealing parameters of 1400°C for 4 h (or a combination of 1400°C for 1 h plus 1350°C for 3 h). The RRR of nine cell resonators after gettering normally reaches values of 500–600 [7, 23].

Post-purification with Ti increases the RRR, while as described above, common annealing at the same temperatures without Ti reduces it (Fig. 8(a) and (b)) [23]. A definite distribution of the interstitial impurities in the cross-section of the cavity wall has to be expected after such purification. The calculated concentration of O, N, and C decreases from the centre to the surface, as shown in Fig. 9 [23]. Similar RRR behaviour is to be expected. A rough method to estimate the RRR distribution of a purified sample was proposed in [23]. A niobium sample after purification heat treatment was etched layer by layer, and the RRR was measured after each step. The RRR distribution of such a sample is shown in Fig. 10. It can clearly be seen that the RRR close to the surface is much higher than that inside the sample, which is in good correlation with the distribution of the interstitial impurities in the cross-section of niobium after post-purification (Fig. 9).

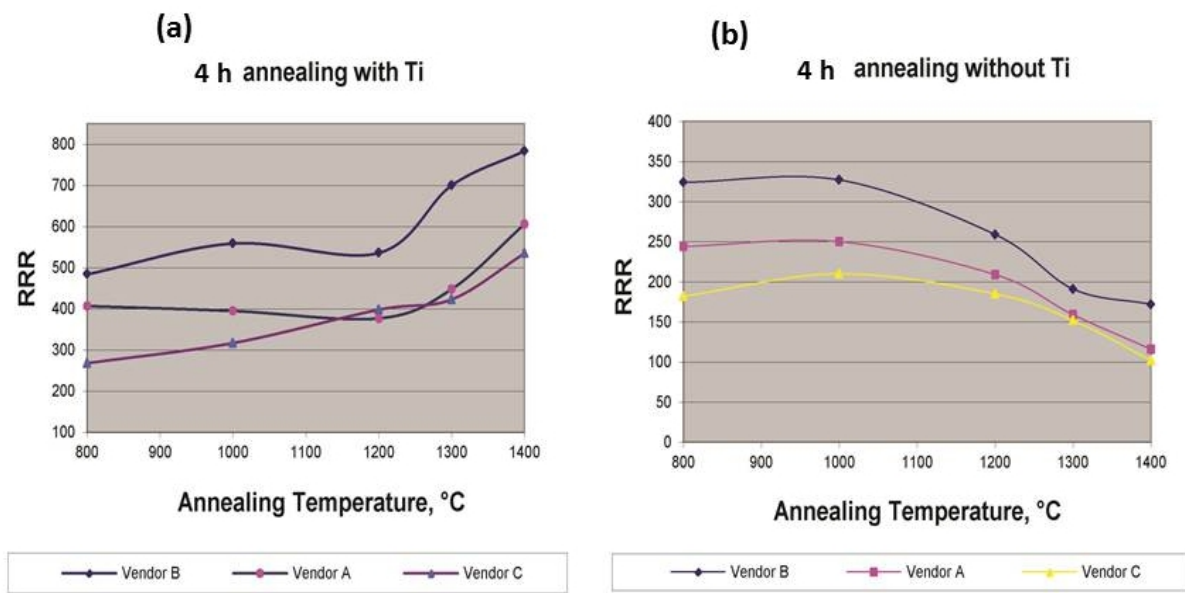

**Fig. 8:** The dependence of RRR on the annealing temperature for Nb annealed (a) with Ti and (b) without Ti

Post-purification potentially allows higher gradients to be achieved. Nevertheless, it has two drawbacks that narrow the range of application of this procedure. First, it reduces the cavity stiffness. The stress–strain curves [7] show that raising the annealing temperature from 800ºC to 1400ºC reduces the elongation at break, the yield and the tensile strength. Secondly, titanium diffuses into niobium during post-purification. It has been shown for FLASH cavities that the Ti diffusion depth is about 60 µm and is deeper in grain boundaries than in grains [7]. This polluted layer of Ti compounds must be removed (as a rule, chemically).

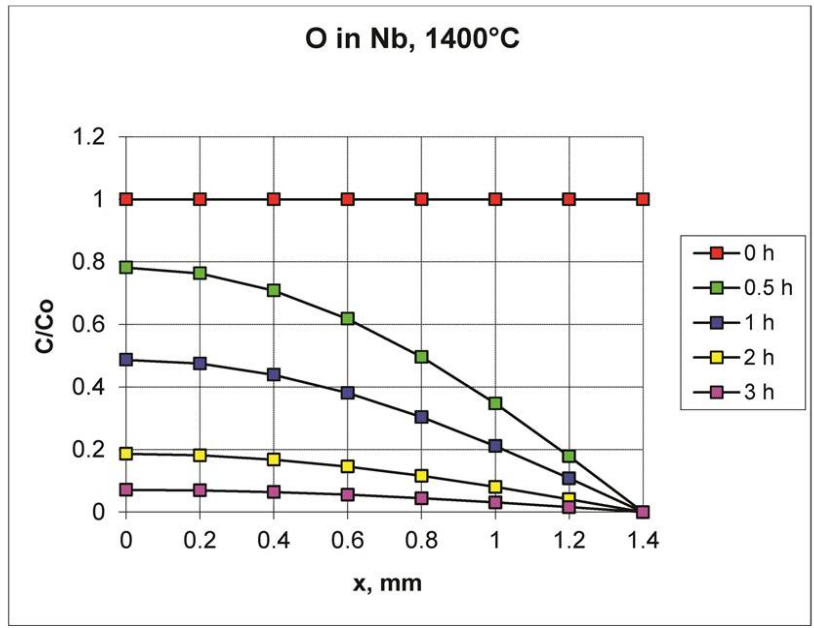

**Fig. 9:** The oxygen distribution from the centre to the surface of an Nb sheet after post-purification

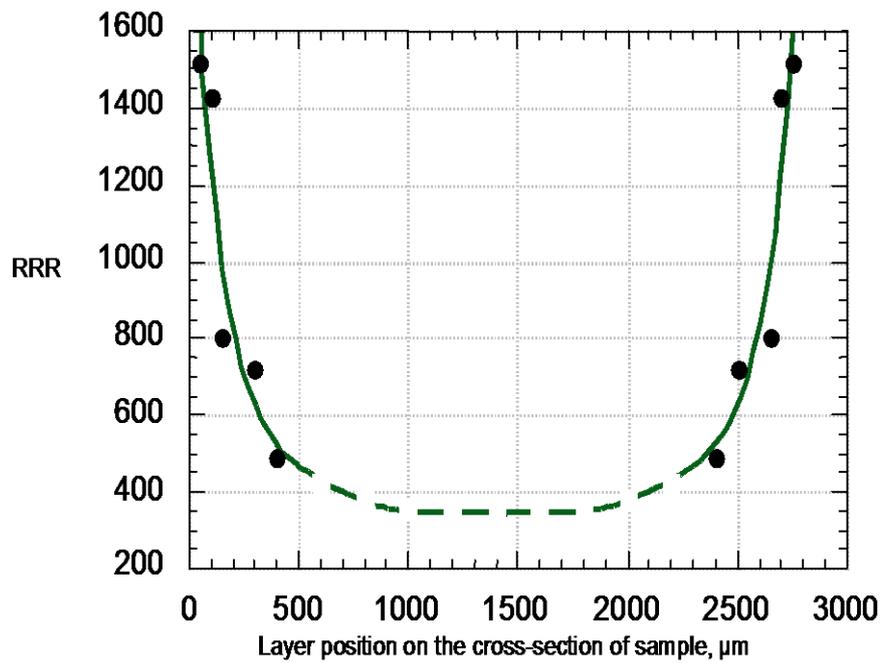

**Fig. 10:** The RRR distribution inside an Nb sheet after post-purification

## 2.4 Specifications of high-purity niobium

The requirements of high-purity niobium for high-gradient superconducting cavities are listed in Table 2.

**Table 2:** Technical specifications for niobium applied for the fabrication of European XFEL 1.3 GHz superconducting cavities [22].

| Electrical and mechanical properties | | The contents of the main impurities in µg/g | | | |
|---|---|---|---|---|---|
| Residual Resistivity Ratio, RRR | > 300 | Ta | ≤ 500 | H | ≤ 2 |
| Grain size | ≈ 50 µm | W | ≤ 50 | O | ≤ 10 |
| Yield strength, $R_p$ 0.2 | $50 < R_p\ 0.2 < 100$ N·mm$^{-2}$ | Mo | ≤ 50 | N | ≤ 10 |
| Tensile strength | >140 N·mm$^{-2}$ | Ti | ≤ 50 | C | ≤ 10 |
| Elongation at fracture | > 30% | Fe | ≤ 30 | | |
| Vickers hardness, HV10 | ≤ 60 | Ni | ≤ 30 | | |

A common indicator of purity, the Residual Resistivity Ratio (RRR), is chosen as RRR > 300.

The main interstitially dissolved impurities that act as scattering centres for unpaired electrons and reduce the RRR and the thermal conductivity are oxygen, nitrogen, hydrogen, and carbon. Oxygen is dominant due to the high affinity of Nb with oxygen. The concentrations of O, N, and C should be kept below 10 µg/g. The influence of hydrogen on the RRR is not so significant, but the hydrogen content should be kept low (less than 2 µg/g) in order to prevent hydride precipitation and degradation of the $Q$-value of the high-RRR cavities under certain cool-down conditions (hydrogen $Q$ decease [1–3, 5]).

Among the metallic impurities, tantalum has the highest concentration (~500 µg/g). As described above, this element accompanies niobium in most ores. An impurity level of 500 µg/g is normally harmless for cavity performance [25], since tantalum is a substitutional impurity and does not substantially affect the behaviour of niobium. Next in abundance among the substitutional impurities are metals such as tungsten, titanium, molybdenum, iron, and nickel, usually at levels less than (30–50) µg/g.

The method of fabrication of Nb sheets at the Tokyo Denkai Company is shown in Fig. 11 as an example [26]. The sheets should be free of defects (foreign material inclusions or cracks and laminations) that could initiate a thermal breakdown. Intermediate and final recrystallization annealing for 1–2 h at 700–800°C in a vacuum furnace at a pressure of $10^{-5}$–$10^{-6}$ mbar has to be performed in order to reach full recrystallization, a uniformly small grain and sufficient mechanical properties for subsequent cavity production (Fig. 12(a)). These conditions can be reached by ensuring a high degree of deformation (> 65%) homogeneously distributed in the bulk of the Nb sheet before annealing [12, 27]. When the deformation is not uniform, striped patterns become visible in the grain structure (Fig. 12(b)). Such sheets have a reduced formability by deep drawing and an increased failure rate.

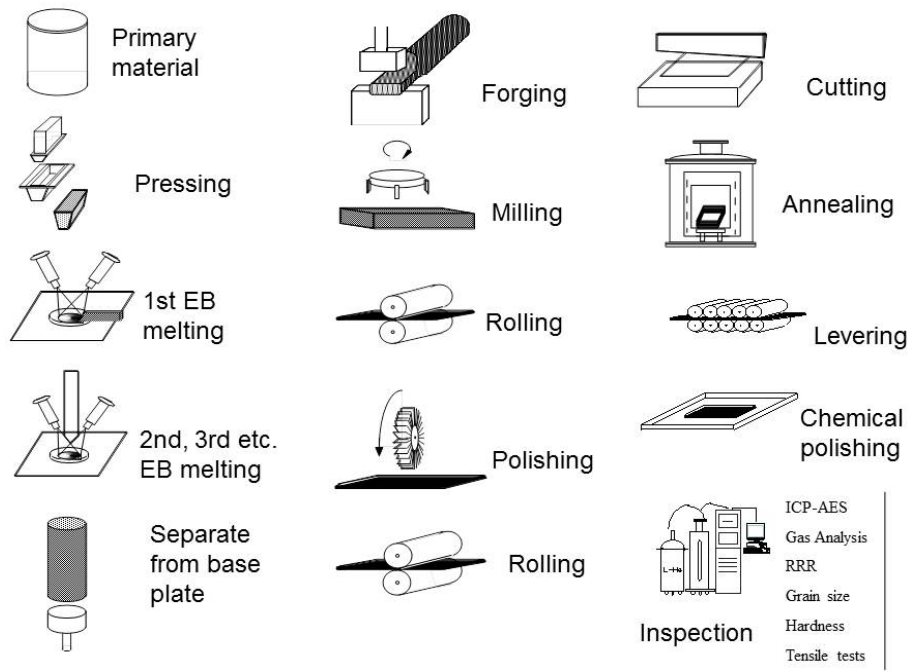

**Fig. 11:** The fabrication of Nb sheets at the Tokyo Denkai Company [26]

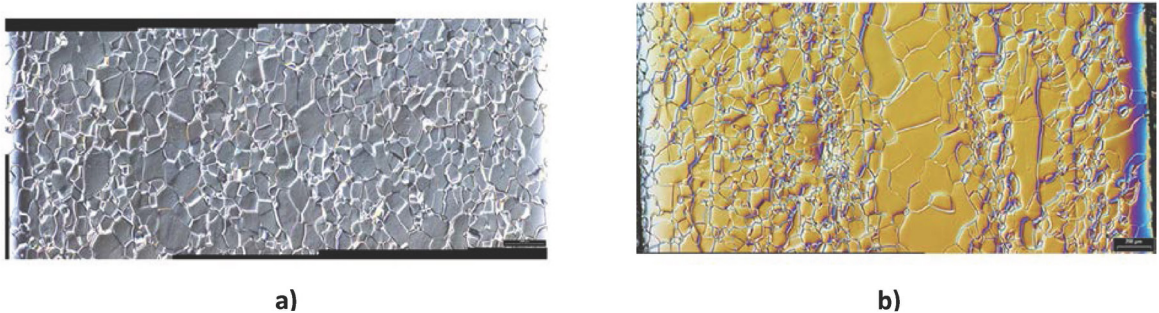

**Fig. 12**: Uniform and homogeneous grains (a) and non-uniform grain structure (b)

Acceptance tests of the material in industry include the RRR measurement, microstructure analysis, analysis of interstitial and metallic impurities, hardness measurement, tensile tests, and examination of surface roughness [22].

Stringent vacuum requirements during EBM and recrystallization heating, and well-controlled conditions during forging, rolling, machining, and grinding, allow the contents of the interstitial impurities in the niobium to be kept at acceptably low levels to ensure the high performance of RF cavities.

## 2.5 Analytics and Quality Control (QC) of niobium

As mentioned above, significant progress in the production of high-purity niobium with an RRR up to 500 has been achieved during the past 20 years on an industrial scale. For example, ~20 tons of Nb with RRR > 300, required for the European XFEL, was produced within 2 years. This progress would be unimaginable without the availability of the appropriate analytical methods and devices.

Three different groups of impurities, easily evaporating metallic elements (substitutional impurities: Al, Fe, Cr, etc.), non-easily evaporating metallic elements (substitutional impurities:

Ta, W), and non-metallic elements (interstitial impurities: H, C, N, O) influence the properties of high-purity niobium. In particular, the content of gases and carbon, and niobium–gas reactions, play a remarkable role during material purification, cavity fabrication, and treatment. Of central importance is the quantitative detection of very low concentrations of gases and carbon at the µg/g level in niobium and the data of their local distribution. A short overview of the analytical determination of these impurities is given below.

*2.5.1   Interstitial and substitutional impurities*

*2.5.1.1   The RRR and thermal conductivity*

The RRR is a common indicator of the level of purity. RRR values do not allow a resolution of the influence of certain elements on electron scattering, but indicate material purity with high sensitivity. This is why most laboratories and companies working with high-purity niobium use the RRR as the first criterion of the level of purity, which can be measured easily and quickly [28].

As is well known from the electron theory of metals, in general cases the electrical resistivity of metals (Mathiessen's rule) at low temperature can be described in the following terms [27]:

$$\rho = \rho_{res} + \rho_{ph}(T) + \rho_{m}, \qquad (1)$$

where the first term is the residual resistivity at $T = 0$ K, caused mainly by electron–impurity scattering and scattering on lattice defects ($\rho_{res} = \rho_{imp} + \rho_{def}$); the second term in Eq. (1) represents the electron–phonon scattering, and $\rho_{m}$ is responsible for the increase of resistivity in a magnetic field.

Scattering of conduction electrons on the lattice (phonon scattering) is absent at $T = 0$ K due to the zero fluctuations of the atoms in the lattice. The $\rho_{def}$ term plays an important role on work-hardened niobium and can reduce the RRR by a factor of up to two. For recrystallized niobium, the $\rho_{def}$ contribution is small. In this case, the total resistivity of Eq. (1) in the absence of a magnetic field consists only of electron–impurity scattering and electron–phonon scattering contributions.

Resistivity caused by scattering of conduction electrons on homogeneously distributed 'chemical' defects (foreign atoms) is proportional to their concentration. The influence of the most important impurity atoms, such as O, N, H, C, Ta, and Zr, on electron–impurity scattering is analysed in some of the reference works [29]. The most popular is the data of Ref. [9], determined on the basis of resistance measurements on niobium voluntarily contaminated by impurities:

$$\rho = \rho_{ph}(T) + \sum \frac{\Delta \rho_i}{\Delta C_i} C_i. \qquad (2)$$

The resistance coefficients $\Delta \rho_i / \Delta C_i$ in Eq. (2) are given in Table 3.

**Table 3:** Residual resistance coefficients of different elements [9]

| Impurity atoms | $\Delta \rho_i / \Delta C_i$ |
|---|---|
| O | 2.64 |
| N | 3.49 |
| C | 3.33 |
| Ta | 0.12 |
| Zr | 0.6 |

In absolutely pure metals that have a lattice without structural defects at temperatures close to 0 K, the resistivity tends to be zero.

The RRR is defined as the following ratio:

$$\mathrm{RRR} = \frac{\rho(300\ \mathrm{K})}{\rho(4.2\ \mathrm{K})},$$

where $\rho(300\ \mathrm{K})$ and $\rho(4.2\ \mathrm{K})$ are the resistivity of Nb at room and liquid helium temperatures, respectively, at standard atmospheric pressure [30]. The superconducting behaviour of Nb (below $T_\mathrm{C} = 9.3$ K, $\rho(4.2\ \mathrm{K}) = 0$) has to be taken into account for RRR determination. The electrical resistivity at 4.2 K is obtained by extrapolation from the value when the material is in a non-superconducting state (Fig. 13) or is obtained due to suppressing the superconducting state of Nb at 4.2 K by application of a high magnetic field. Therefore, in practical implementation, special approaches lead to different measurement methods of RRR for Nb [28, 30].

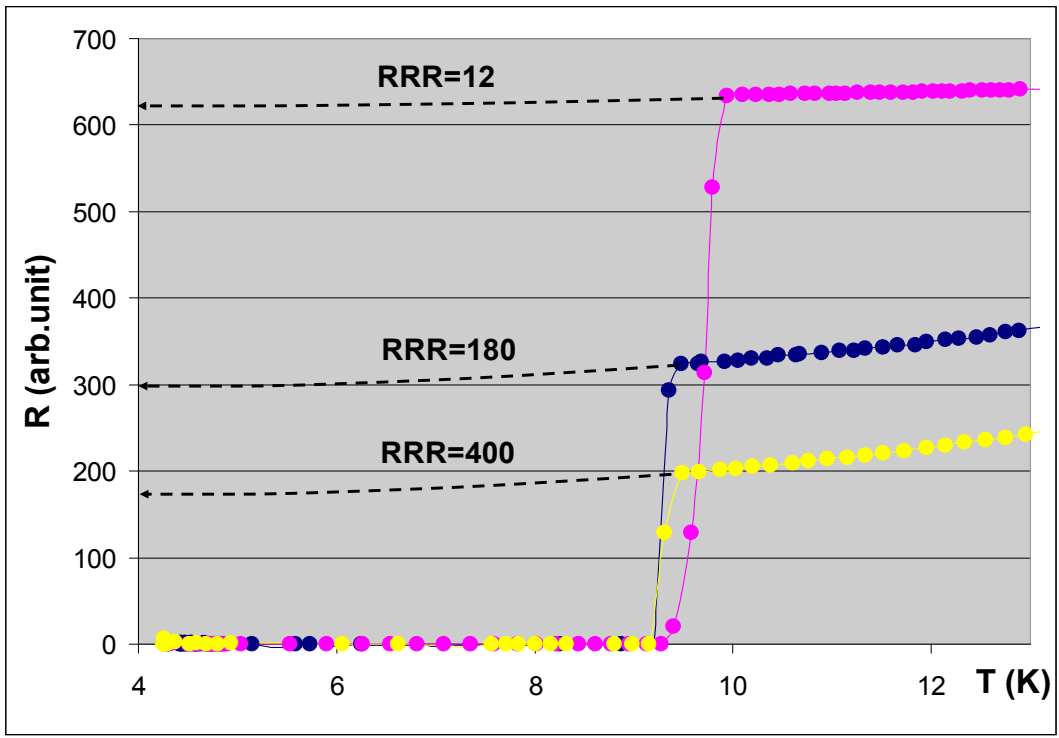

**Fig. 13:** The typical low-temperature behaviour of the electrical resistivity for Nb of different grades of purity

A high thermal conductivity in the cavity wall is needed (at least 10 W·m$^{-1}$·K$^{-1}$ at 2 K) to guide the dissipated Radio-Frequency (RF) power to the liquid helium coolant. For bulk niobium cavities, this requires niobium of high purity [31].

To measure the thermal conductivity, the steady-state method is usually used. The conductivity, $k$, is given by the formula

$$Q = -kA\frac{\mathrm{d}T}{\mathrm{d}x}.$$

By measuring the temperature difference d$T$ over a distance d$x$, with heat input $Q$ and sample cross-section $A$, the thermal conductivity $k$ can be easily derived.

Some curves of the temperature dependence of the thermal conductivity for Nb with RRR between 120 and 760 are shown in Fig. 14. The strong dependence of the thermal conductivity on the purity of the Nb (RRR) can be seen. The rule of thumb giving the simplified relationship between the thermal conductivity and the RRR at liquid helium temperatures is sometimes useful for practical applications [1, 2]:

$$\lambda(4.2\ \text{K}) = C[\text{W}\cdot\text{m}^{-1}\cdot\text{K}^{-1}] \times \text{RRR},$$
$$C \approx 0.25 - 0.14.$$

The following empirical formula can be used to estimate the thermal conductivity of superconducting niobium more precisely, over a wide temperature range [32]:

$$\lambda(T, \text{RRR}, G) = R(y)\left[\frac{\rho_{295\ \text{K}}}{L \times \text{RRR} \times T} + aT^2\right]^{-1} + \left[\frac{1}{D[\exp(y)]T^2} + \frac{1}{BGT^3}\right]^{-1} \ldots \quad (3)$$

The first term in Eq. (3) describes the scattering of electrons by impurities, lattice defects, and phonons; while the second term describes the scattering of phonons by the electrons and the grain boundaries.

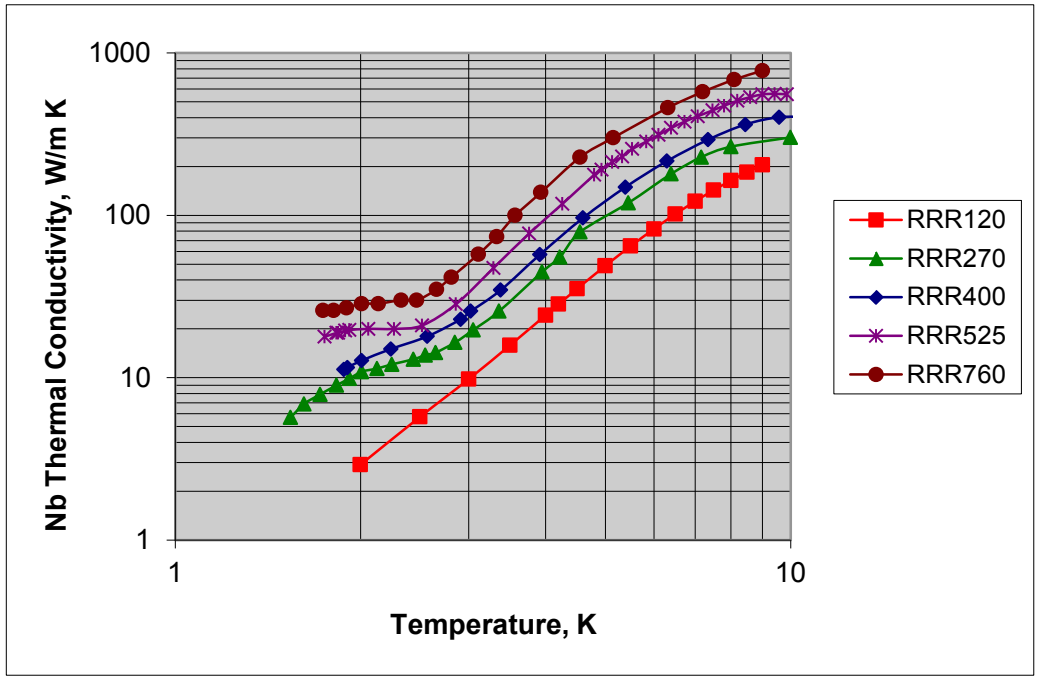

**Fig. 14:** The typical behaviour of the thermal conductivity of polycrystalline niobium at low temperature

The best-fit values for the parameters are to be found in Ref. [32] and are listed in Table 4.

**Table 4:** The parameters for calculation of the thermal conductivity

| Parameter | Best-fit value |
|---|---|
| $L$ | $2.05 \times 10^{-8}$ W·K$^{-2}$ |
| $A$ | $7.52 \times 10^{-7}$ m·W$^{-1}$K$^{-1}$ |
| $\alpha$ | 1.53 |
| $B$ | $4.34 \times 10^{3}$ W·m$^{-2}$K$^{-4}$ |
| $1/D$ | $2.34 \times 10^{2}$ mK$^{3}$·W$^{-1}$ |

*2.5.1.2 Oxygen, nitrogen, carbon, and hydrogen determination*

The classical methods for O, N, C, and H determination consist of extraction of these impurities from niobium in the form of gases, subsequent separation, and determination of quantities. Mostly, vacuum fusion extraction, inert gas fusion extraction, or hot vacuum extraction [10, 33] are used.

In vacuum fusion extraction applied for O, N, and H, a metal sample of defined weight is melted in a vacuum inside a crucible. The crucible is often graphite, so that O is removed as CO. In inert gas fusion, a carrier gas – Ar, for example – is used to sweep out the extracted gas. In hot vacuum extraction, a solid sample is heated in an ultra-high vacuum until the gases completely leave the material.

The released gases are measured by the increase of pressure using a thermal conductivity cell or other pressure-sensitive sensors. In the case of CO, the gas is converted to $CO_2$ by passing over a hot CuO catalyst and the quantity of $CO_2$ is measured by the pressure change or by infra-red absorption. The C content of Nb can be similarly determined in a fusion apparatus, using a flowing stream of $O_2$ to convert C into CO.

For the determination of oxygen, vacuum extraction using the platinum-flux sandwich technique is recommended. An etched sample has to be packed into platinum foil that has previously been degassed in purified helium. The platinum/niobium ratio should be about 10:1.

Industrial devices for O, N, C, and H determination are available on the market: well-known companies are LECO and Horiba. The devices can guarantee 8–10% accuracy [(0.1–0.3) μg/g in absolute values] for hydrogen and 15–25% accuracy [(0.5–2) μg/g] for nitrogen and oxygen.

For the analysis of metallic impurities at levels below (5–50) μg/g, a number of special techniques based on spark mass spectrometry, proton or neutron activation and separation procedures are available. For routine analysis, the main techniques used are Optical Emission Spectrometry (OES), Atomic Absorption Spectrophotometry (AAS), Atomic Emission Spectrometry (AES), Mass Spectroscopy (MS), and X-ray fluorescence analysis.

For strong durable production of high-purity niobium, the industrial laboratory should be equipped with facilities for RRR measurement, scanning apparatus, interstitial impurity analysis (H, N, O, and C), analysis of metallic impurities (Ta, W, and another refractory metals, such as Fe and Ni), metallography, tensile testing, and testing for hardness, HV, and surface quality.

Currently, the companies that are in a position to produce high-purity niobium for superconducting accelerators (five companies worldwide) are mainly equipped with devices for these purposes. For additional investigation that is sometimes required, it is very desirable to have access to analytical equipment for the following: thermal conductivity testing, neutron Activation Analysis (NAA), radiation fluorescence analysis, SEM, EDX, SIMS, XPS, AUGER, X-ray radiography, neutron radiography, texture analysis, and bulge testing (see also Refs. [1–3]).

*2.5.2 Local defects in the sheets for half-cells*

Experiments show that a small spot in just one of the sheets for half-cells can damage the performance of the complete cavity [34]. Thus total quality control of Nb sheets regarding foreign material inclusions and other types of local defects such as cracks, pores, and delamination is very desirable for stable cavity performance.

**Table 5:** Defect diagnostics in Nb sheets

| Method | Principle of the method | Penetration depth | Resolution | Destructive or non-destructive | Test time for a sheet measuring 265 × 265 × 2.8 mm | Remarks |
|---|---|---|---|---|---|---|
| X-Ray radiography | Difference in X-ray absorption between defect and Nb | Complete | Depends on the size of component (10 µm – 1 mm) | Can be non-destructive | About 30 min | The 'shadow picture' depends on the difference in density and atom number |
| Neutron radiography | Difference in neutron absorption between defects and Nb | Complete | Depends on the size of component (10 µm – 1 mm) | Can be non-destructive | About 60 min | The 'shadow picture' depends on the special quality of isotopes, and good detection of light elements |
| Ultrasonic scanning | Reflection of sound waves at interface | Complete (needs a coupling) | Up to 50 µm | Non-destructive | About 30 min | Non-homogeneity in metals is difficult to detect |
| Eddy current scanning | Electromagnetic induction | Depends on frequency (from micrometres to millimetres) | Up to 100 µm | Non-destructive | About 30 min | |
| SQUID scanning | The Josephson effect | Complete | Up to 30 µm | Non-destructive | About 30 min | SQUID scanning devices are not available on the industrial level |
| Neutron activation analysis | Irradiation with thermal neutrons, measurement of γ-spectrum | Complete | Detection of clusters with sizes up to 100 µm | Non-destructive | About 15 h | Efficient for tantalum inclusions, some parts per million of Ta in Nb can be detected |
| Synchrotron fluorescence analysis (SYRFA) | Excitation by white beam (analysis of fluorescence energy spectrum) | 1 µm – 100 µm | Up to 1 µm | Can be done non-destructively | Some hours for inspection of a 20 × 20 mm area | K-lines (energy about 0–80 keV), sensitivity up to a few µg/g of impurity content |
| Synchrotron fluorescence analysis (XAFS) | Energy selection in the primary beam, observation of the absorption edge | About 10 µm | An area of 12 × 12 mm can be tested in one step | Can be done non-destructively | 5 h for inspection of a 150 × 100 mm area | L-lines (energy about 0–10 keV), sensitivity up to a few µg/g of impurity content |
| Microhardness testing | Intrusion of the diamond pyramid | Depends on load value | About 50 µm | Conditionally non-destructive | Residual marks of few micrometres that will be removed during subsequent cavity preparation | Sensitivity depends on the differences in hardness between Nb and inclusion |

The usual conventional surface checks of Nb sheets for cavities are visual inspection, anodization [35] and looking for discolouration, water soaking, and rust traces. It should be taken into account that removal of 100–200 μm thick layers will take place later during cavity preparation, so that inner defects located close to the surface will become uncovered. This means that quality control should be carried out both on the surface and inside the Nb, at a depth of up to 500 μm close to the surface area.

Some important requirements for quality control are as follows.

- It should be non-destructive.
- It should be total. At least one side of the sheet should be scanned. The penetration depth of the signal should not be less than 0.3–0.5 mm.
- It should be fast. The scanning time for one sheet should not exceed 0.5 h.
- It should have a high resolution; defects with a size of 100–500 μm should be detectable.
- It should have sufficient sensitivity to elements with small differences of properties compared to pure Nb (e.g. tantalum).

An overview of the advantages and disadvantages of different quality control methods is given in Table 5.

Up to now, only eddy current scanning has been successfully applied on a large scale for quality control of Nb sheets. Modern eddy current facilities can scan large areas at a rather high speed. High resolution can be achieved by optimizing the electrical and mechanical parameters of the probe. It is possible to detect defects of a size in excess of 100 μm at a depth of a few hundred micrometres. According to simulations of the thermal breakdown (quench), carried out in Ref. [36], a local defect with a size of ~100 μm in high-purity niobium with RRR > 300 caused the quench at $E_{acc}$ to be close to 25 MV·m$^{-1}$. In this case, the requirements of the European XFEL for an accelerating gradient $E_{acc}$ > 23.6 MV·m$^{-1}$ could be fulfilled.

An eddy current scanning device for defect diagnosis of niobium sheets has been successfully developed and applied for the European XFEL at DESY. About 16 000 Nb sheets for the European XFEL have been scanned, and sheets with detected flaws have been sorted out. A typical example of foreign material inclusion (Ta), probably imbedded into an Nb sheet during rolling, can be seen in Fig. 15. Eddy current scanning shows a remarkable local peak, the 3D microscope image indicates a protrusion, and the results of non-destructive element analysis suggested a tantalum inclusion.

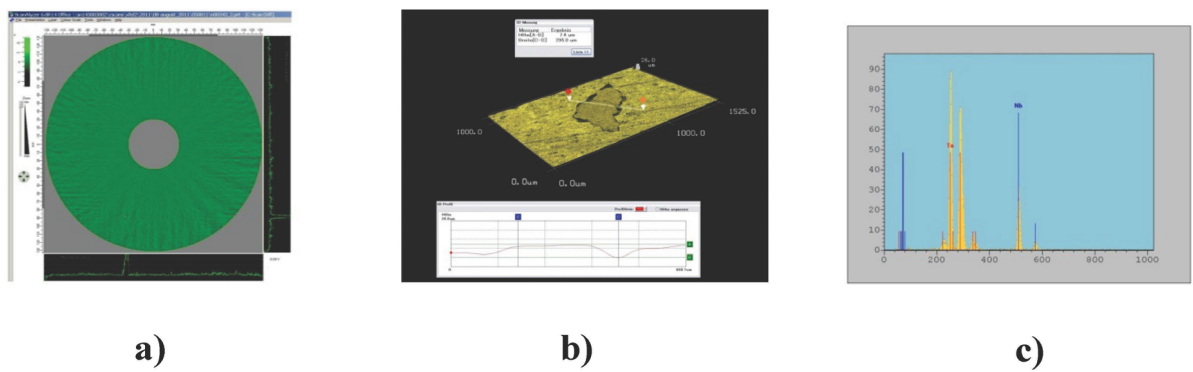

**a)**              **b)**              **c)**

**Fig. 15:** An example of foreign material inclusion (Ta) detected in Nb sheets: (a) eddy current scan; (b) a 3D microscope image; (c) the results of non-destructive element analysis.

Further improvement of the scanning system in order to detect smaller defects in niobium can be done using SQUID-based methods. SQUID sensors are more sensitive in comparison to conventional eddy current pick-up coils [37]. Unfortunately, the industrialization of these devices for this specific purpose has not yet been realized.

## 3 Mechanical fabrication of elliptical cavities

The conventional fabrication procedure consists of the deep drawing and Electron-Beam Welding (EBW) of the parts into a cavity assembly [22]. This procedure is well established and has been used in industrial fabrication for about 30 years.

### 3.1 Deep drawing and EBW

Half-cells are produced from niobium discs, pressed into shape using male and female dies. The dies are usually fabricated from an anodized aluminium alloy. Special hard bronze AMPCO is also suitably applicable to niobium. Deep drawing of half-cells requires high accuracy, because a resonant frequency is generated in this manufacturing step. If the wall thickness deviations in the sheets are less than ±0.1 mm, the use of stiff forming tools only is advisable. Application of an elastic polymer for the female part of the tool allows stable shaping of the inner side of the half-cell independently from the sheet's thickness tolerances. To avoid surface damage, the stamp has to be free from dents, blisters, and scratches.

While establishing the form of the deep drawing tooling, the spring-back of the niobium sheet material has to be taken into consideration. Deep drawing is sensitive to the mechanical properties of niobium. If the grains are too large, an 'orange peeling' roughening effect occurs. If the material is not completely recrystallized, it tears during deep drawing, or deviation from roundness of the half-cell will take place. To achieve good mechanical properties for high-RRR Nb requires the proper choice of annealing temperature and time. The final annealing of 2.8 mm thick Nb sheets normally occurs at 750–800°C in a vacuum oven, at a pressure of $10^{-5}$–$10^{-6}$ mbar for 1–2 h. Niobium also has a low degree of work hardening, which is advantageous for mechanical forming. In almost all cases when the proper mechanical properties of the sheet are achieved, cavity parts can be deep drawn to final shape without intermediate annealing. The accuracy of the half-cell is controlled by tactile 3D measurement and by sandwiching the half-cell between two niobium plates and measuring the resonance frequency.

Two half-cells are joined at the iris with an EB weld, to form a dumb-bell. The iris EBW is usually done partially from the outside and partially from the inside. The next step is the welding of the stiffening ring. Weld shrinkage may lead to a slight deformation of the cell, which needs to be corrected. Afterwards, frequency measurements are made on the dumb-bells to determine the amount of trimming at the equators.

The dumb-bells are visually inspected. Defects and foreign material imprints from previous fabrication steps are removed by grinding. In modern EBW, the iris welds are very smooth, but it is still good practice also to perform the extra step of grinding the iris weld in order to ensure a smooth inner surface in this high-electric-field region and to avoid geometric field enhancement.

Beam tubes are either purchased as extruded seamless tubes or rolled from sheets and EB welded [38]. Flanges for the beam tubes are machined. The flanges are located in regions of low magnetic field and can be made from an Nb–Ti alloy (~55% Ti).

After proper cleaning, eight dumb-bells and two end group sections are assembled in a precise fixture to carry out the equator welding, which is done from the outside. Manufacturers' experience has shown that all equator welds can be done in one action (one evacuation of the EB chamber), without deterioration of the quality.

EBW is usually done in several steps (for more details, see Ref. [22]). The weld parameters are chosen to achieve full penetration. A slightly defocused beam in a circular or elliptic pattern (see Fig. 16), and the use of ~50% of the beam power during the first weld pass and 100% of the beam power in the second pass, allows a smooth weld seam to be obtained. The cavity welding parameters should be adapted to each EBW machine individually. The welding parameters of the DESY EBW machine are listed in Table 6.

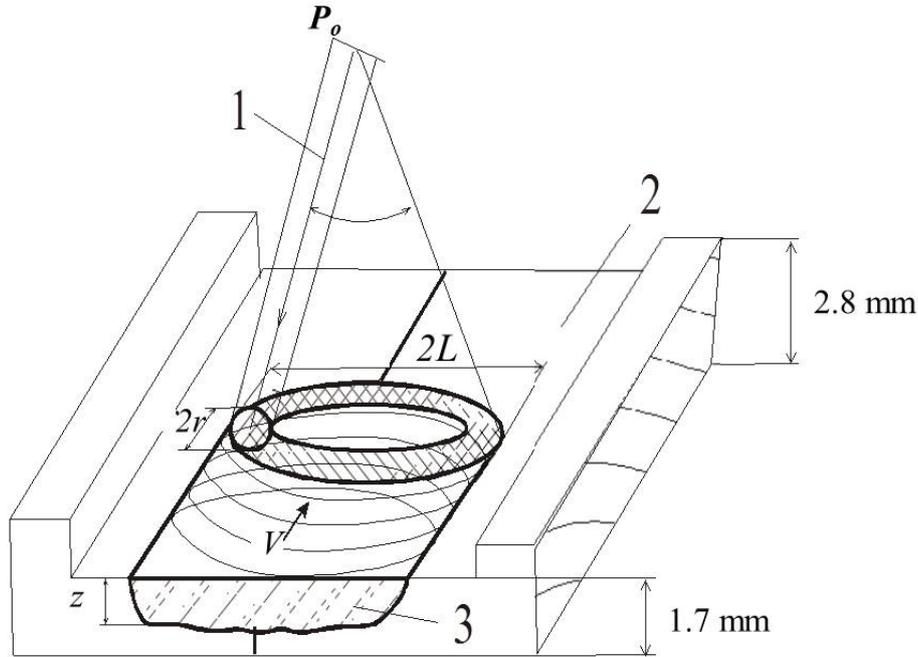

**Fig. 16:** A schematic of the EBW of niobium sheets: 1, the electron beam ($P_0$, power of beam; $r$, radius of slightly defocused beam on the surface; $L$, scanning amplitude; $V$, beam velocity); 2, Nb sheets; 3, melting zone ($z$, depth of the melting zone).

Welding from the inside (RF-side) is recommended, wherever possible. Welds at the equator and iris of cells and at the high-order mode HOM coupler parts are especially critical, because they will be exposed to high magnetic or electric fields. Therefore, thorough cleaning by ultrasonic degreasing, chemical etching, ultrapure water rinsing, and cleanroom drying is mandatory – clean conditions must be assured during welding. Touching the weld preparation area after the last cleaning must be strictly avoided.

The required high quality of the welding seam is in fact very important for performance. This has been demonstrated many times. A direct verification was done, for example, in Ref. [39], due to the investigation of samples separated from the quench areas of cavities with low performance.

Experience shows that sometimes holes can be burned through by EBW. The frequently asked question is whether such a hole is repairable. In Ref. [40], it was shown that the reparation of the burned hole is critical from the point of view of gas contamination. A definite area around the hole is contaminated by oxygen, nitrogen, and hydrogen. It should be removed and replaced by a disc of high-purity Nb before new EBW can be done. Some companies have recently developed a burned hole repair procedure and demonstrated that accelerating gradients up to 30 MV·m$^{-1}$ are reachable in a repaired cavity.

Table 6: Welding parameters for the single-cell cavities of the DESY EBW machine [41]

| Parameter | Iris | | Equator | |
|---|---|---|---|---|
| Beam direction | From inside | From outside | From outside | |
| Maximal duration in ambient air between etching and welding (h) | 8 | 8 | 8 | |
| Percentage of current required for full penetration | – 68 | – 56 | 50 | 100 |
| Angle (degrees) | – 45 | – 0 | 0 | 0 |
| Accelerating voltage (kV) | – 150 | – 150 | 150 | 150 |
| EB current (mA) | – 8 | – 7 | 7 | 13 |
| Beam pattern | – Circle | – Circle | Circle | Circle |
| Frequency (Hz) | – 4000 | – 4000 | 4000 | 4000 |
| Amplitude (mm) | – 1.5 | – 1.5 | 1.5 | 1.5 |
| Focus position set ($\Delta$) | – –25 | – –25 | –45 | –45 |
| Welding velocity (mm·s$^{-1}$) | – 6 | – 6 | 6 | 6 |
| Slope up (angle, degrees) | – 5 | – 10 | 5 | 5 |
| Overlap (angle, degrees) | – 5 | – 5 | 5 | 5 |
| Slop down (angle, degrees) | – 25 | – 30 | 25 | 25 |
| Pressure in working chamber (mbar) | $< 1 \times 10^{-6}$ | | $< 1 \times 10^{-6}$ | |
| Cool-down time (min) | – 120 | – 120 | – | 120 |

### 3.2 Purity degradation during EBW

Electron-beam welding is a rapid process; the temperature goes up and down quickly. The temperature exceeds the melting point of niobium at the welding seam itself, but the temperature in the heat-affected zone depends on the welding parameters, the material thickness, the thermal conductivity of the material, and so on. The process of absorption and desorption of gases during EBW is very complex, and in any case is not in equilibrium. It is in principle possible, but difficult, to make precise calculations of the resulting RRR and the RRR distribution in the welding seam area, but the appropriate data are still lacking. In spite of that, several reasonable experimental results are available.

Since niobium is a strong getter material for interstitial impurities, it is important to carry out the EB welds in a sufficiently good vacuum. The RRR distribution and RRR dependence on pressure in the EB-welded samples from different companies – Dornier, Ettore Zanon, and ACCEL (now RI) – show that the RRR degrades by up to 35% if the total pressure is in the range of $10^{-3}$–$10^{-4}$ mbar. For a pressure between $5 \times 10^{-5}$ and $10^{-5}$ mbar, the RRR degradation is smaller (about 10%) [40]. Tests have shown that RRR300 niobium suffers less than 10% RRR degradation in welding at a pressure lower than $5 \times 10^{-5}$ mbar. Such a vacuum is achievable with industrial equipment and is recommended for the fabrication of superconducting cavities [22]. It is worthwhile to check the RRR degradation of high-purity Nb in the EBW chamber with an ultra-high vacuum (total pressure $< 10^{-5}$ mbar). The investigation in Ref. [40] shows that for the welding seam itself (Fig. 17), the RRR degradation started at a pressure in excess of ~$5 \times 10^{-6}$ mbar. At a pressure below $5 \times 10^{-6}$ mbar, the RRR in the welding seam even improves, from ~350 to 370–380. The improvement of the RRR is maximal for the pressure region from $10^{-8}$ to $5 \times 10^{-7}$ mbar. It is interesting that the improvement in the RRR almost does not depend on the pressure in this region. This means that the effort to reach a vacuum better than $5 \times 10^{-7}$ mbar does make sense, at least for the above-mentioned welding parameters.

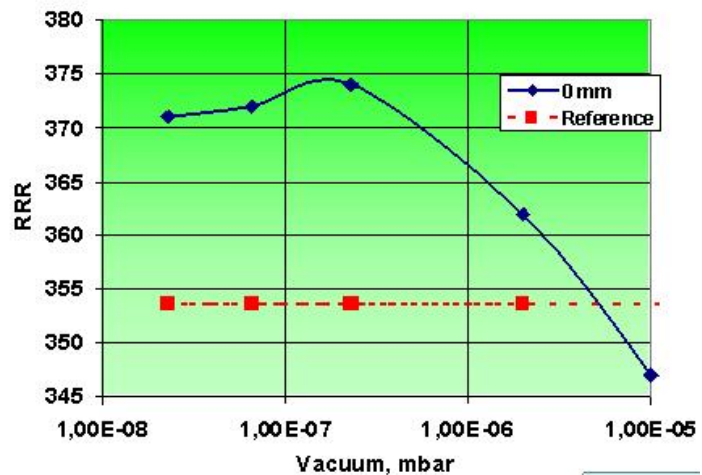

**Fig. 17:** The RRR in the welding seam versus pressure

On the other hand, it is easy to realize that RRR degradation takes place in the heat-affected area (see Fig. 18). The RRR degradation is maximal in the region ~10–15 mm away from the welding seam (from 350 to 320–330).

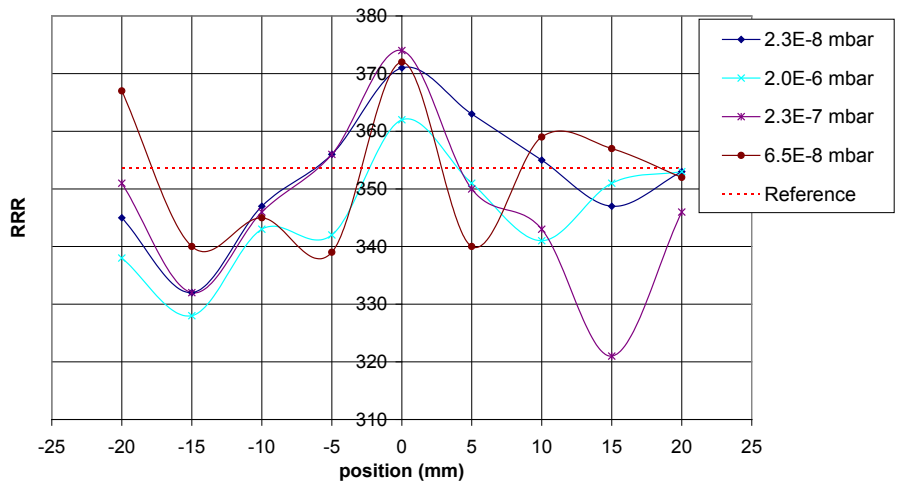

**Fig. 18:** The RRR for the EBW area versus distance from the welding seam at different pressures (DESY EBW facility).

### 3.3 The RRR and the hydrogen distribution in the weld area

The hydrogen content measured by heat extraction in the welding seam as well as in the heat-affected zone [40] shows a good correlation with the RRR, as can be seen in Fig. 19. Reduction of the RRR follows with enhancement of the hydrogen content. Evidently, hydrogen absorption does not take place directly in the welding seam, but in the areas with rather moderate temperatures (the heat-affected zone). It is well known, from pressure-concentration isotherms of hydrogen for Nb in a steady-state condition (Fig. 20), that at temperatures above 500 K and under the common partial pressures of hydrogen, the hydrogen content in Nb becomes less than 1 μg/g. At lower temperatures, absorption of hydrogen will take place according to Fig. 20. Obviously, the heat-affected area fulfils these last conditions, which then result in an enhanced hydrogen content.

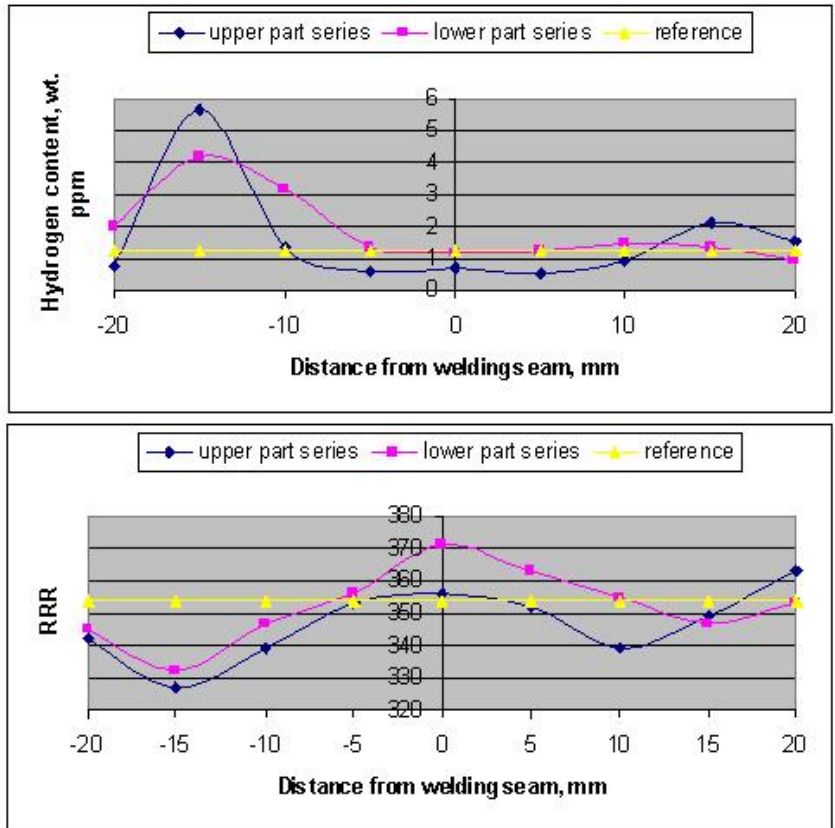

**Fig. 19:** The RRR distribution compared with the hydrogen content in the weld area (pressure $2.3 \times 10^{-8}$ mbar)

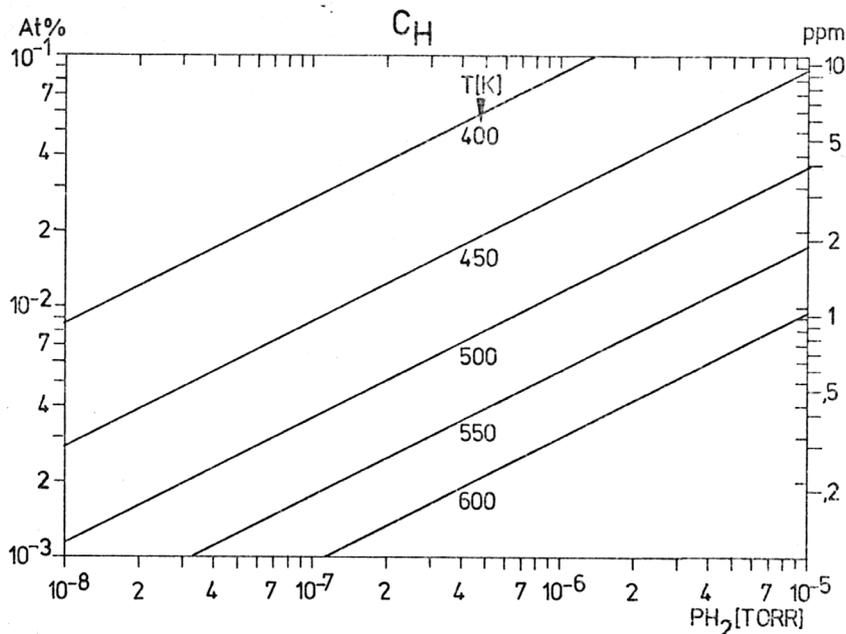

**Fig. 20:** Pressure-concentration isotherms for hydrogen in Nb in the steady-state condition [10]

An enhanced hydrogen content at the weld area is in good agreement with the results of Refs. [42, 43], where the hydrogen disease of Nb cavities was explored. It was pointed out that the weld area is susceptible to hydrogen disease even if the chemical treatment of the cavity was carried out correctly. This illustrates why the annealing of cavities at 800°C during treatment (hydrogen degassing) is reasonable.

An example of Rest Gas Analysis (RGA) during welding is shown in Fig. 21 [44]. The start of welding is normally accompanied by a significant rise of the hydrogen partial pressure in the chamber. The partial pressure of the water decreases at the same time, as can be seen in Fig. 21. Probably, the electron beam brings about the decomposition of water molecules, which produces a lot of hydrogen.

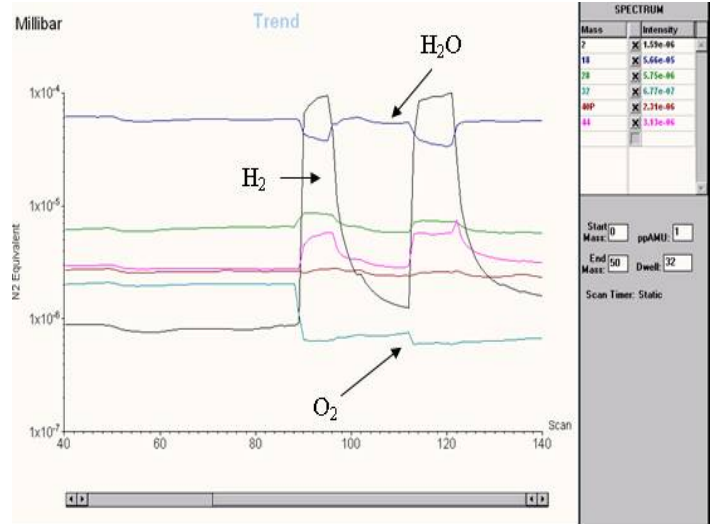

**Fig. 21:** An example of the partial pressure in the EB chamber during the welding of a Nb 300 sample

### 3.4 Cavity length adjustment procedure

The main principle of the length adjustment procedure follows Ref. [45]. Uncertainty in the shape of half-cells has to be corrected before completion of the final cavity. The correction should fulfil the aim that the cavity finally has the right frequency and the right length. The procedure takes advantage of the fact that the frequency change due to a shape change is different for regions with a high magnetic field (equator) and a high electric field (iris). A trimming of the half-cell near the equator ($\Delta L_e$) increases the frequency; a compression of the half-cell in the iris region ($\Delta L_z$) decreases the frequency. The frequency-sensitivity factors are listed in Table 7 [45].

**Table 7:** Sensitivity factors

|  | Equator, $\Delta L_e$ | Length, $\Delta L_z$ |
|---|---|---|
| $\delta f$ [MHz·mm$^{-1}$] | −5.3 | 5.4 |

The length adjustment method consists of the following steps.

1. Frequency measurement of the half-cells.
2. Dumb-bell assembly (welding of iris and stiffening rings).
3. Frequency measurement of the dumb-bell.
4. Trimming of the dumb-bell at the equator, if necessary.
5. Positioning of the dumb-bell in the cavity.
6. Cavity production (equator welding of dumb-bells and end groups).
7. Frequency and length measurement of the cavity.
8. Calculation of the expected cavity length after tuning.

The individual steps of the length adjustment procedure for the half-cell, dumb-bell, end group, and complete cavity are described in detail in the specification for the production of European XFEL cavities [22]. Descriptions of the quality-assurance and other documentation can also be found in those documents.

## 3.5 Quality management and documentation

Some main principles of quality management that have to be kept in mind during cavity production are mentioned below [22, 46].

The contractor (the cavity manufacturer) and all its subcontractors have to create and maintain a Quality Management (QM) system according to DIN ISO 9001. Their quality control system has to be independent of the manufacturer. The list of subcontractors must include evidence that each subcontractor has a certified QM system according to DIN ISO 9001, or must contain information on how quality is assured and audited there. The orderer (the institution placing the contract for cavity production) may check a subcontractor's qualifications by carrying out quality audits and holding adjustment and determination discussions.

The orderer normally requests some quality checks including protocolling in addition to the Quality Assurance/Quality Control (QA/QC) measures carried out by the contractor. Protocols must be reported at a time close to the period of component and cavity fabrication.

For each completed cavity, a traceability report has to be prepared, in which the serial numbers of all incorporated semi-finished products and manufacturing groups are stated in the form of a parts list, including position information. A conclusion concerning the raw material used should be possible.

The manufacturer has to prepare a Conformity Certificate (CC, confirmation of conformity) for each completed cavity. The CC confirms that the cavity has been produced according to specification and that all requirements have been checked and that they comply with the orderer's demands. A Non-Conformity Report (NCR) must be prepared if the properties of a certain component deviate from specification. In the NCR, the additional procedure has to be proposed by the contractor. The NCR has to be accepted by the orderer.

The experience of European XFEL cavity production shows that use of the Engineering Data Management System (EDMS) as a central repository for all engineering information and for the complete documentation relating to the cavity fabrication process, and prompt exchange with the manufacturer, are advisable [47, 48] (see Fig. 22). This implies that all documents (inspection sheets, specifications, drawings, etc.) that are created and maintained during the RF cavity manufacturing process should be made available electronically in the EDMS. The exact formats and implementation of the test protocols have to be agreed on between the contractor and the orderer. The complete information that must be reviewed by the manufacturer or approved by the orderer shall be controlled by the system. Cavity producers have access only to relevant documents and data. For statistical analysis, a database system, as is operational at DESY [49], is very convenient.

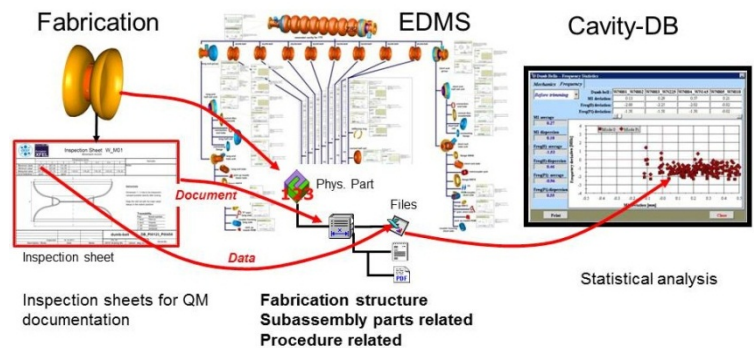

**Fig. 22:** An example of cavity data management using EDMS and a data bank

### 3.6 Initial experiences in the production of superconducting cavities for the European XFEL [50]

The production of superconducting cavities for the European XFEL includes the following: mechanical fabrication, the use of material provided by DESY, prior EP treatment, ethanol rinsing, outside etching, 800°C annealing, tuning to resonant frequency, final surface treatment by Buffered Chemical Polishing (BCP) or EP, High-Pressure Water (HPR) rinsing, 120°C baking, integration of the cavity into a Helium Tank (HT), assembly of the HOM, pick-up and high-$Q$ antennas, and shipment to DESY for the vertical RF test.

The fabrication of 800 serial cavities, the largest in the history of cavity production, is going on at two companies – Research Instruments (RI) in Germany and Ettore Zanon in Italy – on the 'build to print' principle and is planned to be finished in 2015.

A new infrastructure, mostly for cavity treatment, is being created at both companies. The infrastructure comprises EBW equipment, ISO 7 and ISO 4 cleanrooms with cleaning, rinsing, and etching facilities, Ultra-Pure Water (UPW) production systems, HPR rinsing, 800°C annealing furnaces, tools for cavity integration in the HT, a 120°C final baking oven, a Slow-Pumping Slow-Venting (SPSV) vacuum system, and systems for visual inspection of the cavity's internal surface.

DESY provided both companies with an in-house developed sophisticated machine for cavity tuning at room temperature (Cavity Tuning Machine (CTM)) and equipment for RF measurement of dumb-bells and end groups, called HAZEMEMA.

DESY and colleagues from INFN/LASA are monitoring the production process. The main principle of the supervision of the production is that the cavities have to be built strictly according to the XFEL specifications, but a performance guarantee is not required. The supervision consists of: a quality control plan (also for the Pressure Equipment Directive, or PED – see below); the internal QA and QM systems of the companies; NCRs; and regular visits to the company by DESY expert teams. In addition, there are regular 'Project Meetings' at the company location (approximately every month, depending on the production progress and quality).

Information flow from the companies to DESY/INFN is organized to be paperless, as described above, utilizing the EDMS. Transfer of documents from data systems to EDMS is fully automated.

Another important issue for European XFEL cavity production is the implementation of the Pressure Equipment Directive (PED). The fabrication experiences of the European XFEL cavity with a helium tank as a pressure-bearing component according to European requirements are briefly described below.

To avoid a pressure test on a complete cryo-module with eight cavities prepared for electron acceleration and dressed with power couplers and other accessories, the following options (modules) B and F are chosen for cavity testing according to European Directive PED/97/23/EC.

Module B (EC type examination) includes: examination of the design; FEM calculations; qualification of welding processes; qualification of other PED-relevant processes (annealing, deep drawing, forming); production and destructive examination of test pieces; supervision of production qualifications on the first eight series cavities; and PED-relevant testing methods for series production of the cavities.

Module F (product verification) is dedicated to series production and includes mainly visual inspections, monitoring of documents, and a pressure test for each cavity. The contracted 'notified body' (TUEV NORD) advises and tracks the process.

An important stage of this activity was the qualification of the test pieces. A test piece is composed of two cells with a helium vessel, without end groups, representing all pressure-bearing parts and welds of the cavity. It is built using exactly the same manufacturing methods and welding parameters that are used in the series production. Two test pieces per company were produced and successfully qualified by destructive examination.

Semi-finished products (cavity material) for pressure-bearing sub-components of cavities and helium tanks also have to be qualified and purchased, according to PED 97/23/EC, by companies qualified for this task. These PED activities on material for cavities and helium tanks consist of qualification of cavity materials Nb40, Nb300, Nb - Ti, Ti Grade 1, and Ti Grade 2 itself (creation of the Particular Material Appraisal, or PMA); certification of the QM system at the companies producing the cavity material and the sub-components for the helium tanks; and supervision of the procurement of the semi-finished material products (traceability, marking, 3.1 test certificates, etc.).

Meanwhile, the first hundreds of series cavities have been produced and treated using the newly qualified infrastructure and shipped to DESY for cold RF testing. Most cavities shipped up to the present time have immediately satisfied the XFEL specifications.

## 4 Large-grain and single-crystal niobium

### 4.1 Introduction

The idea of using ingot material for cavity fabrication in the form of sheets sliced from an ingot and producing cavities by means of deep drawing and EBW technology was introduced at Jefferson Lab (JLab). It turned out that the sliced material had sufficiently good mechanical properties and could be formed into half-cells despite the large grains. This manufacturing approach, for Large-Grain (LG) cavities, has attracted worldwide interest in the past few years [51–54]. This option allows the long production chain from large-grain ingot to fine-grain sheet to be eliminated, and seems to be more cost-effective than the conventional fine-grain method. A lot of studies on material characterization, cavity fabrications, and testing around the world have appeared since that time. An overview of these investigations can be found in Ref. [51].

### 4.2 Material, fabrication, and RF performance

Up to now, industrial slicing of the discs has been developed by two companies: W.C. Heraeus in Germany and the Tokyo Denkai Company in Japan [55–57].

One of the issues to be solved was efficient cutting of the discs. W.C. Heraeus have developed a wire saw cutting procedure that can be cost-effective, assuming that the company can carry out the slicing of many tens of discs simultaneously while keeping the material purity high (RRR > 300), and achieving tight thickness tolerances (better than ±0.1 mm) and a high surface quality ($R_a$ < 1.6). The principle and the main part of the W.C. Heraeus machine for cutting of the large-grain ingot by wire sawing can be seen in Fig. 23. The procedure developed at Tokyo Denkai is similar [56].

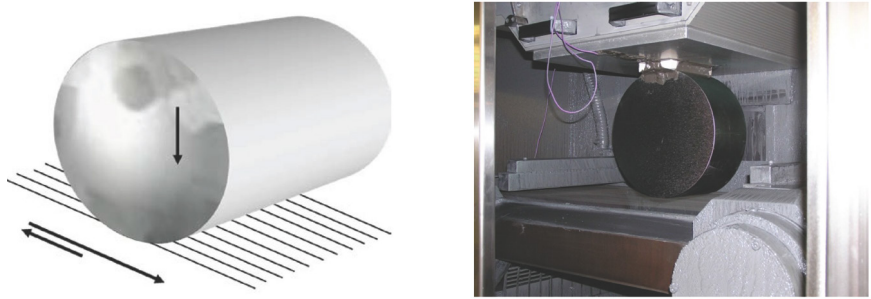

**Fig. 23:** The principle and the machine for cutting of the large-grain ingots by wire sawing at W.C. Heraeus. (Courtesy of W.C. Heraeus.)

It can be expected that disc material from ingots will be less vulnerable to contamination by foreign material or other types of defects, because it is taken directly from homogeneous ingot material that has been very slowly re-melted many times. Several types of flaws (delamination, imbedded particles, oxides, etc.) that can occur during forging or rolling are avoided in this case. Experience has shown that the eddy current scanning usually applied to fine-grain sheets is useless [52, 55].

The introduction of ingot niobium for SRF cavities has attracted the interest of many institutions. Work on fabrication and testing of such cavities has been carried out at Jefferson Lab, DESY, KEK, Michigan State University, Cornell University, Peking University, the Institute of High Energy Physics of China, and BARC of India.

Investigations with single-cell cavities have shown that a rather high accelerating gradient $E_{acc}$ of usually 25–35 MV·m$^{-1}$ can be achieved by rather simple BCP treatment with HF(40%):HNO$_3$(65%):H$_3$PO$_4$(85%) in the volume ratio 1:1:2. Especially comprehensive investigations have been undertaken at JLab on single-cell cavities produced with material from different companies, of different LG dimensions, with different tantalum contents (300–1200 µg/g), and different shapes and resonant frequencies (for more details, see Ref. [51]). It has been shown that even using BCP treatment, an accelerating gradient of 45 MV·m$^{-1}$ is achievable. For high performance, approximately 100 µm of material has to be removed from the surface when conventional procedures such as BCP and hydrogen degassing are used.

In contrast to most laboratories, which have made use of their own fabrication and rapid prototyping capabilities for exploring ingot niobium, DESY has collaborated closely with industry. After encouraging results gained on the single-cell LG cavities, the main aim was to analyse the potential of ingot material for large-scale applications such as the European XFEL. Eleven nine-cell LG cavities of TESLA shape were produced at ACCEL Instruments GmbH – now Research Instrument GmbH (RI) – from W.C. Heraeus material.

It was shown that it is feasible to build not only single-cell, but also nine-cell, cavities from LG material without significant difficulties. Deep drawing of the half-cells was done using the same tools as for fine-grain material. The grain boundaries were noticeably pronounced, with steps of up to 0.5 mm. The deep-drawn half-cells had a quadrangular or oval shape, and sometimes did not meet the required tolerance of ±0.2 mm. It turned out that the presence of large central crystals in the discs and their crystallographic orientation had a big influence on the shaping. It is well known that main atom plane slipping for Body-Centred Cubic (BCC) metals takes place in the (110) planes in the [111] direction; therefore, for the discs with (100) orientation, a more pronounced anisotropy and quadrangular shape after deep drawing were expected, in agreement with observations.

The pronounced shape deviation in the half-cells generated some difficulties for assembly of half-cells, and especially of dumb-bells for welding. RI overcomes these difficulties by using a special tool that ensures precise joint assembly of the male and female half-cells. Application of the DESY length adjustment procedure [45] allowed the correct cavity length and the required frequency of the fundamental mode to be achieved in all of the nine-cell cavities.

After the BCP treatment (removal of a 100 µm surface layer inside and a 20 µm layer outside, annealing at 800°C for 2 h, final BCP of 20 µm inside, followed by baking at 125–135°C for 48 h), accelerating gradients up to 30 MV·m$^{-1}$ in π-mode measurement ($B_p$ = 110–130 mT) and up to 35 MV·m$^{-1}$ in pass band measurement have been achieved in a stable and reproducible manner for all 11 nine-cell LG TESLA-shaped cavities [52].

After additional EP consisting of main EP of about 50–70 µm, removal followed by an ethanol rinse, an additional standard 800°C firing, and a final EP consisting of ~50 µm removal, ethanol rinse, six final HPR cycles, and baking at 120°C, the performance of the cavities was significantly improved, up to 31–45.5 MV·m$^{-1}$ at $Q_0$ values above 10$^{10}$, limited mostly by breakdown. Figure 24 shows the plot of the unloaded quality factor versus the acceleration gradient $Q_0(E_{acc})$ at 2 K for these cavities. Enhancement of the acceleration gradient, typically by more than 10 MV·m$^{-1}$ after EP in comparison to BCP was observed.

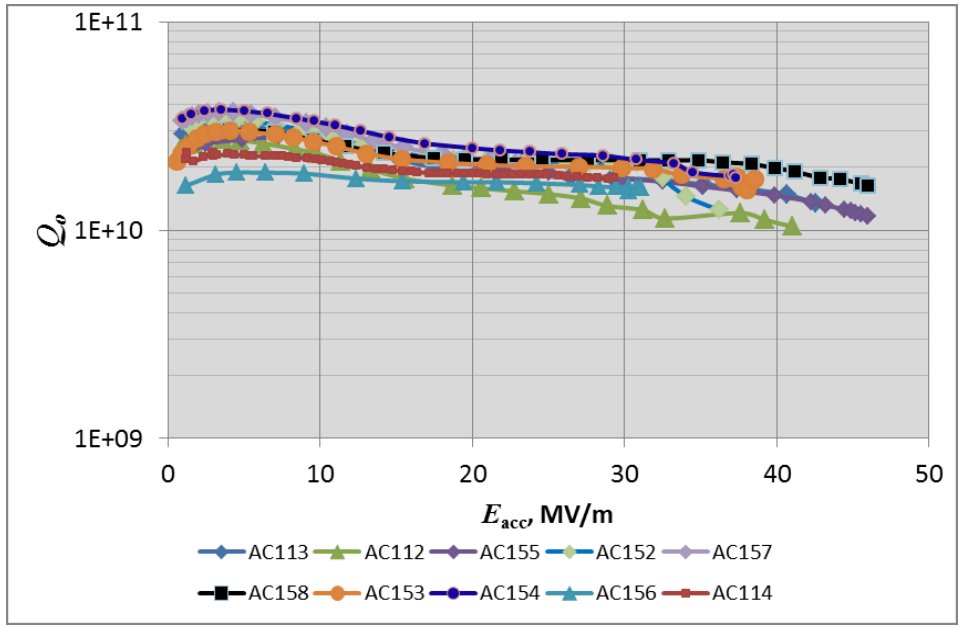

**Fig. 24:** The final $Q_0(E_{acc})$ performance of the LG cavities AC112-AC114, and AC151-AC158 at 2 K after EP

Another interesting aspect of LG cavity behaviour that needs to be stressed is the rather high unloaded quality factor, $Q_0$, after EP treatment, even at high gradients. As can be seen in Fig. 25, $Q_0$ approaches $3.5 \times 10^{10}$ at 2 K, at moderate accelerating gradients. Figure 25 compares the $Q_0$ values of 11 EP-treated LG cavities with the $Q_0$ values of 15 XFEL prototype cavities (AC115–AC129) treated according to an XFEL recipe [46]. As can be clearly seen, the $Q_0$ value for the LG cavities is ~25–30% larger than for conventional fine-grain cavities, which indicates the high potential of LG cavities in applications requiring high $Q_0$ values; for example, in Continuous Wave (CW) applications. The superiority of the $Q_0$ value of LG cavities after BCP treatment is less pronounced.

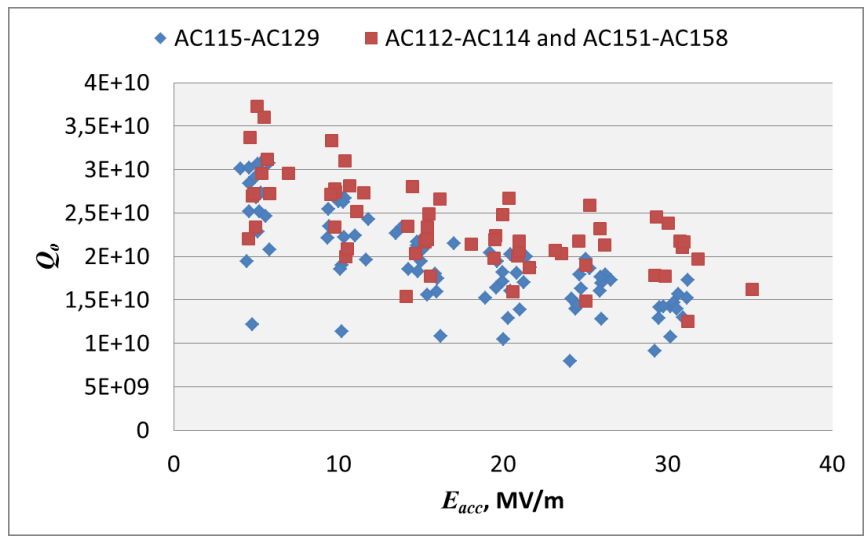

**Fig. 25:** A comparison of the unloaded quality factor $Q_0$ at 2 K for 11 EP-treated LG cavities (red) with $Q_0$ at 2 K for XFEL prototype cavities (AC115–AC129, blue, best result) treated according to an XFEL recipe (partly Final EP and partly BCP Flash [24] were applied).

The complete chain of the LG cavity technique, beginning with material production and ending with cavity installation into a cryo-module, has been successfully tested at DESY. Two LG cavities are installed and operational in the FLASH accelerator at DESY at the present time. The first European XFEL cryo-module made from LG cavities has been produced and tested, and it fulfils the specifications requirements.

**4.3　Advantages and disadvantages of LG cavities**

Based on experience and on discussions of the past few years at workshops and conferences [51–54, 58], the advantages (pros) and disadvantages (cons) of LG cavities are summarized below.

*Pros*

1. LG discs are more cost-effective than fine-grain sheets (by 32%, according to the estimate for XFEL pre-series cavities).
2. The wire saw procedure allows us to achieve high surface quality and thickness tolerances in the discs.
3. The increased thermal conductivity close to 2 K due to the phonon effect helps us to lead the heat away from hot spots.
4. Simplified quality control is possible. There is no danger that during the many production steps from ingot to sheet, the material will be polluted (no RRR degradation). Eddy current scanning is avoidable.
5. An accelerating gradient of 25–30 MV·m$^{-1}$ can be reached by simple preparation with BCP only. The best result of 45 MV·m$^{-1}$, reached after EP, is one of the best results for this type of cavity.
6. The onset of $Q$-drop in large-grain cavities is typically at 10% higher accelerating gradients.
7. It is sufficient to bake the BCP-treated cavity at 120°C for only 12–24 h.
8. The complete chain of the LG cavity technique, beginning with material production and ending with cavity installation into a linear accelerator, is proven.
9. The quality factor $Q_0$ of EP-treated LG cavities is ~25–30% larger compared to similar fine-grain cavities.
10. The wire saw method causes much less stress at the surface of the disc compared to rolled sheets (it reduces the influence of the damage layer on performance).

*Cons*

1. LG is currently not usable for mass production. For example, the industry is not in a position to produce the required amount of ~20 tons of LG material for the European XFEL in 2–3 years.
2. Only one company has industrial experience of LG cavity production.

**4.4　Single-crystal cavities**

Why does EP treatment allow the attainment of better performance for LG cavities than BCP? The surface quality of large grains is comparable for BCP and EP treatment. The surface roughness of the BCP-treated large grains depends on the crystal orientation, but is on the same level as EP-treated fine-grain material (hundreds of nanometres [52], and see Fig. 26).

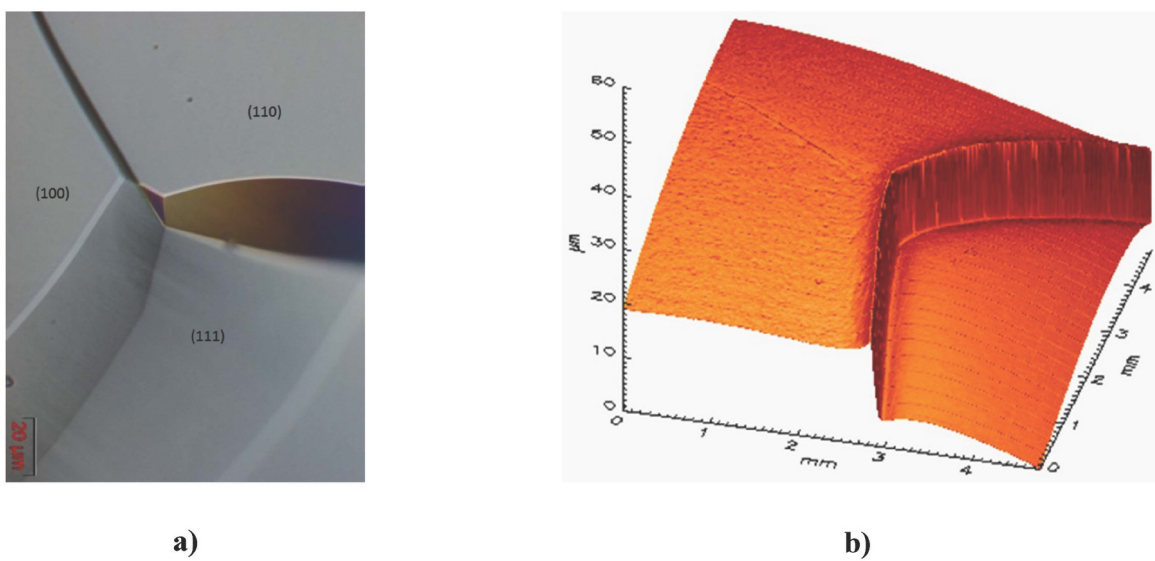

**a)** **b)**

**Fig. 26:** (a) A light microscope image of the grain boundary triple junction of LG Nb after 100 μm BCP on the previously mechanically polished sample: the orientation type is shown on the grains. (b) An atom force microscope image of the same area: the steps on the grain boundaries are within of 1.5–15 μm for this orientation constellation.

It seems that the difference in performance between BCP and EP treatment of LG cavities is caused by the Grain Boundaries (GBs). The reduced $E_{acc}$ of BCP-treated cavities can be explained, for example, by the magnetic field enhancement at grain boundary edges and earlier penetration of the external magnetic field in the niobium, which has a geometrical nature [1, 2]. In other words, GBs can be considered as planar weak links with a reduced critical current density. In addition, GBs enhance the possibility of hydrogen absorption and diffusion in the areas in which impurities gather [1–3].

From this point of view, it would be reasonable to take into consideration the behaviour of the single-crystal cavities (cavities without grain boundaries). A single-crystal cavity with no grain boundaries definitely has the potential to improve cavity performance and simplify the treatment procedure substantially, because by nature it does not contain any interruption of the crystal lattice orientation.

Is it possible to produce cavities consisting of a one single crystal? Indeed, a fabrication method for single-crystal cavities has been proposed and a few single-cell single-crystal cavities have been produced recently [59, 60]. The following aspects have been demonstrated on samples and taken into consideration for the fabrication proposal (for more details, see Refs. [59, 60]).

- Definite enlargement of the single-crystal disc diameter is possible without destroying the single-crystal structure.
- The single crystals retain the crystallographic structure and, after shaping of the cavity half-cell from a disc by deep drawing, the orientation perpendicular to the surface remains.
- Appropriate heat treatment will not destroy the deformed single crystal.
- If the orientation of the crystals is matched, two single crystals will grow together by EBW.

Especially important for cavity fabrication by EBW are the last points, which allow production of cavities as a complete single crystal. It has turned out that two single crystals will grow into one single crystal if the crystallographic orientations are matched at the EBW seam with an accuracy approaching 3°. The results of metallographic analyses of cross-sections of niobium samples can be seen in Fig. 27(a) and (b). Whereas unmatched orientations produce a pronounced grain boundary

(Fig. 27(a)), matched orientations of both single crystals grow together without an interface (grain boundary free) (Fig. 27(b)).

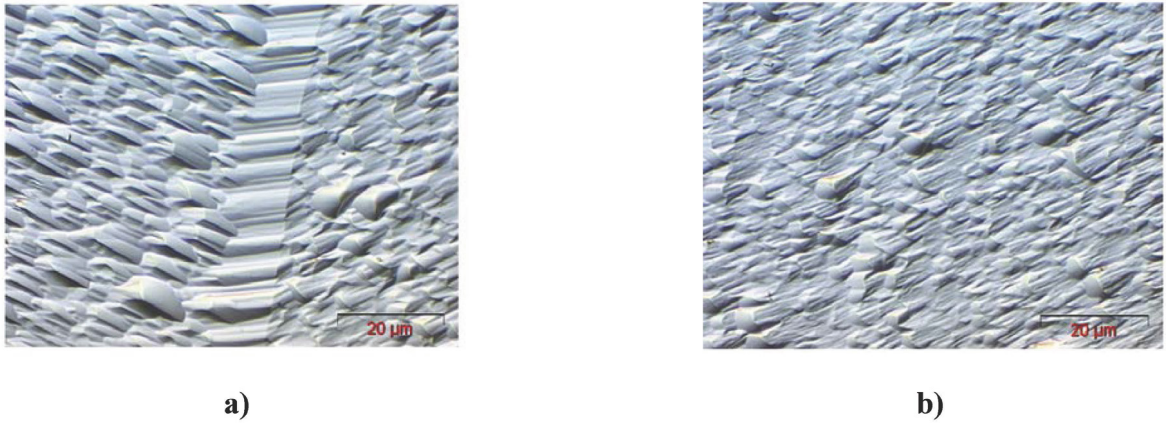

| a) | b) |

**Fig. 27:** (a) The EBW connection of two single crystals without matching of crystal orientation (the grain boundary is pronounced). (b) The EBW connection of two single crystals after matching their orientations.

Several single-crystal cavities have been produced at JLab (of 2.3 and 2.8 GHz resonant frequency, without enlargement of a single crystal) and at DESY (1.3 GHz, by applying the enlargement procedure). It has been demonstrated that, for BCP-treated single-crystal cavities, a similar performance as that after EP treatment of LG cavities, namely an accelerating gradient at the level of 40 MV·m$^{-1}$, is reachable. The main results for the performance of single-crystal cavities are summarized in Table 8.

**Table 8:** A summary of the test results for the single-crystal single-cell cavities produced at JLab and DESY

| Cavity number and resonant frequency | $E_{acc, max}$ (MV·m$^{-1}$) | $B_{peak, max}$ (mT) | $Q_0(B_{peak,max})$ | Treatment |
|---|---|---|---|---|
| 1 (2.3 GHz) | 38 | 162 | $4.0 \times 10^9$ | 200 µm BCP, 800°C/3 h, HPR, 120°C/48 h |
| 2 (2.3 GHz) | 45 | 160 | $7.0 \times 10^9$ | 200 µm BCP, 800°C/3 h, HPR, 120°C/24 h |
| 1AC6 (1.3 GHz | 41 | 177 | $1.2 \times 10^{10}$ | 250 µm BCP, 750°C/2 h, 120 µm EP, HPR, 135°C/12 h |
| 1AC8 (1.3 GHz) | 38.9 | 168 | $1.8 \times 10^{10}$ | 216 µm BCP, 600°C/10 h, HPR, 120°C/12 h |
| 5 (2.8 GHz) | 38.5 | 166 | $7.6 \times 10^9$ | 170 µm BCP, HPR 120°C/12 h |

A single crystal applied up to now for cavity fabrication had a crystal orientation close to (100). In this context, the question arises of how cavity performance can be influenced by crystal orientation. This issue has not been investigated until now and is definitely relevant. Anisotropy of the superconducting energetic gap $2\Delta/T_c$ in the 4d and 5d transition metals, including Nb, is well known [61]. As one of the main parameters of superconductivity, $\Delta$ should influence the main physical properties of the material (e.g. the critical magnetic field $H_c$ and the surface resistance). Therefore, it is not unimaginable that the fabrication of single-crystal cavities with a preferred orientation could lead to further improvement of performance.

On the other hand, the enlargement procedure for a single-crystal disc is costly. Several attempts have been made in the industry to produce, in a stable manner, rather large single-crystal discs with diameters of 200–300 mm. Generally, the procedure for single-crystal creation is similar to the well-known vertical Bridgman procedures for single-crystal growth; that is, partially melted seed, an axial temperature gradient, and a movable interface between the solid and liquid phases are applicable. Nevertheless, the industry is still not in a position to produce such single crystals of high-purity niobium.

## 5    Fabrication of seamless cavities

The idea of fabricating seamless elliptical cavities is attractive. It is to be expected that this technology could not only reduce the production costs, but also improve the cavity's accelerating performance. Some efforts have been made in the past through the application of explosive forming [62], spinning [63], and hydroforming [64]. Explosive bonding was successfully applied to Cu tubes, but failed for bulk niobium. Spinning was started in the 1990s at INFN Linearo, and up to now it has been demonstrated that a single-cell spun cavity can reach an $E_{acc}$ similar to that for a welded cavity up to 40 MV·m$^{-1}$.

Most efforts in recent years have been focused on hydroforming, and this technique has been applied to elliptical cavities from 300 MHz to 3.9 GHz in many laboratories: CERN, Cornell in the 1980s, CEA SACLAY in the 1990s, and KEK and MSU (Michigan State University). Many laboratories now successfully produce cavities in copper. A hydroformed 1.3 GHz single-cell cavity from CEA SACLAY, made from Nb of RRR = 20, has reached $E_{acc}$ = 18 MV·m$^{-1}$ after post-purification [65].

In the 1990s, DESY started and established a highly consistent hydroforming programme for TESLA-like cavities [66–68]. The DESY forming procedure consists of two stages: reduction of the tube diameter in the iris area and subsequent expansion of the tube by hydroforming. One starts with a tube of intermediate diameter between iris and equator. The hydroforming expansion generally consists of three steps: determination of the stress–strain properties of the tube material, computer simulation of the expansion, and the hydroforming itself.

The hydraulic two-dimensional bulging of the disc into the spherical form was used to produce a stress–strain diagram. The values of the stress, $\sigma$, and the strain, $\varepsilon$, are determined by measurement of the pressure, $p$, the radius of curvature, $r$, and the thickness, $t$, in the zenith of the sample during the deformation procedure. The numerical simulation of the hydraulic expansion of the tube was done at DESY using the finite element code ANSYS [68]. Non-linear elasto-plastic behaviour in accordance with the stress–strain curve and isotropic hardening rules were taken into account. The calculations were carried out on the basis of the experimentally determined stress–strain characteristic of the tubes to be hydroformed, and resulted in a relationship between the applied internal pressure versus the axial displacement with radius growth (the path of the expansion) for the hydroforming process.

It was found that the strain before necking of niobium can be increased by using a periodic stress fluctuation (pulse regime). Tensile tests have shown that elongation before necking is almost 30% higher by application of a pulse method in comparison with a monotonous stress increase. In addition, the strain rate during the hydraulic expansion of Nb should also be taken into account. Experiments have shown that the deformation procedure should be rather slow. Around 10% of strain before necking can be achieved by keeping the strain rate below $10^{-3}$ s$^{-1}$.

During the actual forming process, an internal pressure is applied to the tube and simultaneously an axial displacement, forming the tube into an external mould; deformation is controlled by a PC, which follows the established simulation protocol. The main criterion for hydraulic expansion is the theoretically determined relationship between the internal pressure and the axial displacement. Further correction of the expansion parameters can be done on basis of comparison of the theoretical and experimental growth of the tube diameter.

The development of the production of seamless Nb tubes for hydroforming was done in collaboration with a number of companies. Tube production [69] by spinning, and back extrusion in combination with flow forming and deep drawing, were checked. A uniform, small-grain, and homogeneous texture is required to provide high plastic deformation of the tube during hydroforming. The tubes produced from the Nb sheet (spinning or deep drawing, with subsequent flow forming) show higher strain before the onset of necking compared to those produced by back extrusion combined with flow forming. Several single-cell units of 1.3 GHz TESLA-shape niobium cavities have been produced at DESY. Hydroformed single cells have reached an accelerating gradient $E_{acc}$ approaching 35 MV·m$^{-1}$ after BCP, and approaching 42 MV·m$^{-1}$ after EP (for an example, see Fig. 28).

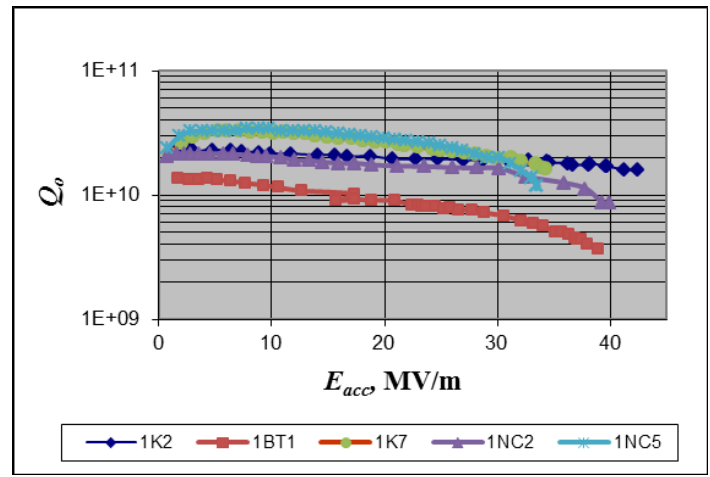

**Fig. 28:** The $E_{acc}$ performance of hydroformed single-cell cavities

The work of the past few years has been concentrated on multicell and nine-cell cavities. Several seamless two- and three-cell units have been produced. An accelerating gradient $E_{acc}$ of 30–35 MV·m$^{-1}$ has been reached after BCP, and an $E_{acc}$ approaching 40 MV·m$^{-1}$ has been reached after EP. Hydroformed three-cell units combined with nine-cell niobium cavities have been completed at Ettore Zanon. The accelerating gradient reached was $E_{acc}$ = 30–35 MV·m$^{-1}$ (Fig. 29). One cavity has been successfully integrated into a cryo-module and is operational in the FLASH accelerator at DESY.

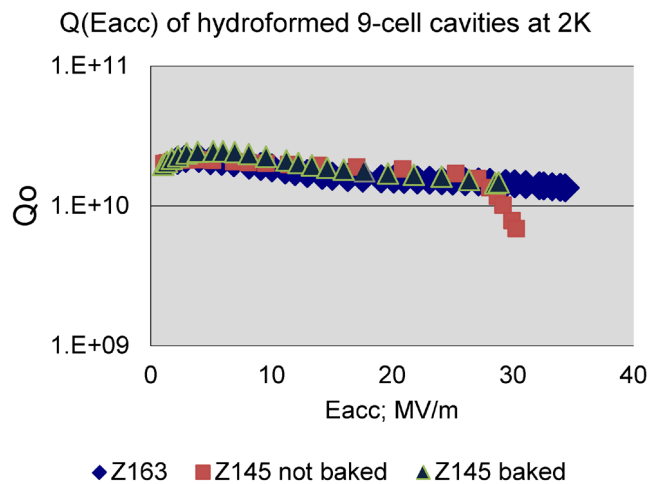

**Fig. 29:** The $Q_0(E_{acc})$ performance of hydroformed nine-cell TESLA-shape cavities